\documentclass[12pt]{article}
\usepackage{a4wide}
\usepackage{latexsym}
\usepackage{amsmath}
\usepackage{amsfonts}
\usepackage{amscd}
\usepackage{amsthm}
\usepackage{amssymb}
\usepackage{cite}
\usepackage{shuffle}
\usepackage{hyperref}

\usepackage{pslatex}
\usepackage{graphicx}
\usepackage[latin1,utf8]{inputenc}
\usepackage[T1]{fontenc}
\usepackage{euscript}
\usepackage{subfigure}
\usepackage{bm}

\usepackage{mathtools}
\usepackage{arydshln}

\newcommand{\bq}{\begin{eqnarray}}
\newcommand{\eq}{\end{eqnarray}}

\theoremstyle{plain}

\usepackage{xcolor}

\allowdisplaybreaks

\begin{document}

\vspace{1.5cm}

\begin{center}
 {\Large\bf Analytic result of a three-loop integral family \\
 in the Higgs decay to four massive bottom quarks\\
  }

  \vspace{1cm}
  
 {\large Jian Wang${}^{a,b}$, Xing Wang${}^{c,d}$, Yefan Wang${}^{e,f}$\\
 \vspace{1cm}
     {\small \em ${}^{a}$  School of Physics, Shandong University, Jinan, Shandong 250100, China}\\
     {\small \em ${}^{b}$ Center for High Energy Physics, Peking University, Beijing 100871, China}\\     

     {\small \em ${}^{c}$  School of Science and Engineering, The Chinese University of Hong Kong, Shenzhen, } \\
     {\small \em Longgang, Shenzhen, Guangdong 518172, China}\\
     {\small \em ${}^{d}$  Southern Center for Nuclear-Science Theory (SCNT), Institute of Modern Physics, } \\
     {\small \em Chinese Academy of Sciences, Huizhou, Guangdong 516000, China}\\
     
     {\small \em ${}^{e}$  Department of Physics and Institute of Theoretical Physics, Nanjing Normal University,}\\ {\small \em  Nanjing, Jiangsu 210023, China} \\
     {\small \em ${}^{f}$  Nanjing Key Laboratory of Particle Physics and Astrophysics, Nanjing Normal University,}\\{\small \em  Nanjing, Jiangsu 210023, China}

 } 
\end{center}


\begin{abstract}\noindent
When calculating the decay width of $H\to b\bar{b}b\bar{b}$ induced by the effective Higgs-gluon-gluon interaction using the optical theorem,  the three-loop Feynman diagrams are encountered. Among them, there is an interesting integral family, which involves not only the sectors evaluated to multiple polylogarithms but also those related to an elliptic curve and a K3 surface. 
We demonstrate how to construct and solve the canonical differential equation for the entire family, which includes mixing between different sectors, 
by using the recently proposed algorithm in~\cite{e-collaboration:2025frv, Bree:2025tug}. 
Analytical results for all master integrals in this family up to the first two orders of $\varepsilon$ are given. 

\end{abstract}

\newpage
\tableofcontents

\section{Introduction}

\label{sec:introduction}The decay into a bottom-anti-bottom quark pair channel accounts for the largest fraction of the Standard Model Higgs decays and provides one of the cleanest handles on the bottom quark Yukawa coupling. 
The HL-LHC program and the proposed Higgs factories aim at percent-level, and eventually sub-percent-level, determinations of this coupling \cite{ATLAS:2018jlh,Zhu:2022lzv,CEPCPhysicsStudyGroup:2022uwl,Altmann:2025feg}. This experimental prospect makes the high-precision theory prediction indispensable. The perturbative description of the Higgs decay to bottom quarks induced by 
the bottom quark Yukawa coupling has been refined for decades, from the early QCD corrections \cite{Braaten:1980yq,Sakai:1980fa}  to more recent higher-order calculations in the massless-bottom approximation \cite{Chetyrkin:1995pd,Baikov:2005rw,Herzog:2017dtz,Davies:2017xsp}. Although the terms induced by the top Yukawa coupling begin to contribute at formally higher perturbative orders, their numerical impact is significant and thus can never be neglected. 
Since the top quark is much heavier than the Higgs and bottom quark, it proves convenient to integrate out the top quark, resulting in the effective Lagrangian contains a gluonic operator $O_1$ and a bottom Yukawa operator $O_2$. 
Their Wilson coefficients, denoted by $C_1$ and $C_2$, respectively,  feed directly into the decay rate.  The interference proportional to $C_1C_2$ begins at order $\alpha_s^2$ and generates logarithms of $m_H^2/m_b^2$, whereas the $C_1C_1$ contribution starts one order later, but it is power-enhanced compared to the pure bottom-Yukawa term.  
These effects have been studied at fixed orders in several recent works \cite{Primo:2018zby,Mondini:2020uyy,Wang:2024ilc}, and the $C_1C_1$ contribution with massive bottom quarks at order $\alpha_s^4$ was obtained in \cite{Wang:2026zfy}.  The present paper is a follow-up to that calculation.  Rather than focusing on the inclusive decay width and the phenomenological impact,  we solve analytically a full integral family underlying the calculation.  Besides the results of this highly non-trivial three-loop integral family, which are beyond multiple polylogarithms (MPLs),   the systematic approach to tackle this problem is worthy of emphasis  since it  can be applied in other processes.

The modern approach to calculating general multi-loop Feynman diagrams rests on two closely related pillars. On the one hand, integration-by-parts (IBP) identities \cite{Tkachov:1981wb,Chetyrkin:1981qh} reduce all integrals in a given integral family to a finite set of master integrals. For integrals relevant for cutting-edge precision predictions, IBP reduction relies heavily on computing resources despite using efficient public packages~\cite{Lee:2013mka,vonManteuffel:2012np,Maierhofer:2017gsa,Smirnov:2019qkx,Klappert:2020nbg,Peraro:2019svx,Wu:2023upw,Guan:2024byi}. Some packages are based on (variants of) the Laporta algorithm \cite{Laporta:2000dsw}, while some packages also take advantage of syzygy relations, module intersection, and finite-field reconstruction methods. Recently, new ideas have been proposed; see, for example, \cite{Wang:2024hsm, Feng:2025leo,delaCruz:2026mas,Smith:2025xes,Dlapa:2026oyq,Feng:2026imq}. In particular, the reference \cite{Feng:2026imq} gave a very good overview of IBP reduction improvements and a summary of recent progress in different directions. On the other hand, one needs to evaluate the master integrals after IBP reduction. The most efficient and general method for this purpose is the differential equation \cite{Kotikov:1991pm,Remiddi:1997ny,Gehrmann:1999as} (together with IBP reduction). There have been a lot of advances during the last two decades. On the numeric side, some efficient public packages based on differential equations were developed recently, e.g., \cite{Hidding:2020ytt,Liu:2017jxz,Liu:2022chg,Armadillo:2022ugh,Chen:2024xwt,Prisco:2025wqs,Liu:2026cpf,Czakon:2026tog}. On the analytic side, it was observed that a well-chosen basis of master integrals can make the dimensional regulator $\varepsilon$ stand out as a prefactor of the connection matrix and thus transform the differential equation from a coupled computational problem into an iterated description of the function space~\cite{Henn:2013pwa}. Such a well-chosen basis is usually called the canonical basis, in particular in the context of multiple-polylogarithmic (MPL) situations. Then, the $\epsilon$ expansion unfolds in terms of Chen iterated integrals \cite{Chen:1977oja}. In many cases, the resulting functions are polylogarithms, harmonic polylogarithms (HPL)~\cite{Remiddi:1999ew}, and their generalization,  MPLs~\cite{Goncharov:1998kja}. 
The understanding of rich mathematical structures of MPLs, e.g., symbols and the Hopf algebra \cite{Goncharov:2010jf,Duhr:2011zq,Duhr:2012fh}, has helped to push the theoretical predictions at colliders to unprecedented precision during the last decade. The simplicity of the canonical basis evaluated in MPLs is closely tied to integrands with logarithmic singularities and constant leading singularities \cite{Arkani-Hamed:2010pyv,Henn:2013pwa}, 
and there exist algorithms and public packages to derive such a basis, e.g., \cite{Lee:2014ioa,Prausa:2017ltv,Gituliar:2017vzm,Meyer:2017joq,Lee:2020zfb,Dlapa:2020cwj,Henn:2020lye}. Starting already at two loops, however, physically important Feynman integrals go beyond MPLs. The equal-mass sunrise graph represents the first example \cite{Laporta:2004rb,Bloch:2013tra,Adams:2013nia}, and the study of the associated elliptic structure has led to new functions and concepts such as elliptic polylogarithms and iterated elliptic kernels
\cite{Adams:2015gva,Adams:2016xah,Remiddi:2017har,Broedel:2017kkb,Broedel:2018iwv,Broedel:2018qkq}. Interested readers can refer to \cite{Weinzierl:2022eaz} for a comprehensive review. 

Progress in the last decade has made it clear that elliptic curves are just the beginning of the story beyond MPLs, e.g., \cite{Adams:2018yfj,Delto:2023kqv,Marzucca:2025eak,Becchetti:2025oyb,Schwanemann:2024kbg}.  Multi-loop banana graphs furnish a particularly systematic sequence of Calabi-Yau examples
\cite{Klemm:2019dbm, Bonisch:2020qmm}, with the three-loop equal-mass banana linked to a K3 surface \cite{Primo:2017ipr,Pogel:2022yat} and the four-loop banana to a Calabi-Yau threefold \cite{Pogel:2022ken,Pogel:2022vat,Pogel:2022yat}.  Calabi-Yau and higher-genus structures also arise in multi-loop scattering amplitudes, fishnet integrals, self-energy integrals, and gravitational scattering \cite{Bourjaily:2018ycu,Bourjaily:2018yfy,Duhr:2022dxb,Duhr:2022pch, Duhr:2024bzt, Forner:2024ojj,Driesse:2024feo,Duhr:2025lbz,Bargiela:2025vwl,Brammer:2025rqo,Lellouch:2026xdr,Duhr:2026apb,Yang:2026rgb}. These developments have shifted the focus from solving isolated examples to understanding the function spaces, period matrices, and monodromy structures that organize the whole classes of Feynman integrals.
At the same time, the notion of an $\varepsilon$-factorized or canonical basis has been broadened radically, and the previous algorithms for solutions in terms of MPLs are no longer applicable in general. Several prescriptions have been proposed to generate $\varepsilon$-factorized differential equations for integral families with elliptic curves or geometries of higher complexity
\cite{Pogel:2022yat,Pogel:2022ken,Pogel:2022vat,Duhr:2022dxb,Gorges:2023zgv,Delto:2023kqv,Duhr:2024bzt,Duhr:2024uid,Frellesvig:2024rea}.
Gradually, it is realized that the $\varepsilon$-factorized basis and the resulting function space are closely related to the underlying geometry. 
This deepens the connection between high-energy physics and mathematics. 
Most recently, an algorithmic framework for the construction of a $\varepsilon$-factorized basis, proposed by the $\varepsilon$-collaboration~\cite{e-collaboration:2025frv,Bree:2025tug},
can deal with Feynman integrals corresponding to different geometries in a unified way. 
The essential idea is that Feynman integrals not only have block structures, usually called sectors, but also enjoy multi-layered sub-structures inside each sector, which decompose the vector space spanned by master integrals in a given sector into smaller subspaces, similar to filtrations in Hodge theory. Apart from various examples with the corresponding maximal cuts given in \cite{e-collaboration:2025frv,Bree:2025tug}, this algorithmic method has been adopted to obtain the $\varepsilon$-factorized basis of the whole three-loop banana integral family with four different masses \cite{Pogel:2025bca}. 
However, this problem does not have non-trivial mixing between different sectors.

In this work, we apply and generalize the algorithm to the integral
family contributing to the $H\to b\bar{b}b\bar{b}$ decay, which belongs to the real correction to the $H\to b\bar{b}$ decay at order $\mathcal{O}(\alpha_s^4)$.
This integral family can be considered as a three-loop problem after using the optical theorem, as shown in Fig.~\ref{fig:higgs3L}. 
Under the four massive cuts, it has four sectors and features an elliptic curve in the top sector, a K3 surface related to the three-loop equal-mass banana 
as the lowest sector, and the other two sub-sectors of the MPL type. 
In this companion article to \cite{Wang:2026zfy}, we focus on deriving the $\varepsilon$-factorized differential equation for all the above sectors in an algorithmic and unified way based on recent papers~\cite{e-collaboration:2025frv,Bree:2025tug} and investigate the solutions in detail. In particular, we consider mixing between different sectors and investigate the solution space to demonstrate how the algorithm is applied in an integral family instead of a sector alone. We believe that the detailed description in this manuscript can help to understand the content of \cite{Bree:2025tug} 
and thus is useful for the calculations of other Feynman integrals.

The paper is organized as follows.  In section~\ref{sec:setup}, we define the integral family, summarize the two steps of the algorithm, and set the stage. We apply the strategy of ``separate and conquer'' to calculate the diagonal blocks one by one in section~\ref{sec:diag}. 
After that, we deal with the remaining mixing among sectors in section~\ref{sec:mixing}.  Section~\ref{sec:de} presents the $\varepsilon$-factorized differential equations, and section~\ref{sec:bd} gives the required boundary conditions. They lead to analytic results of the master integrals, shown in section~\ref{sec:sol}.  We conclude in section~\ref{sec:conclusions}.

\section{Setup}
\label{sec:setup}
The integral family is defined as:
\begin{equation}
    \label{eq:intdef}
	\begin{aligned}
		I_{\nu_1\cdots \nu_9} = e^{3\varepsilon \gamma_E } \left(m^2_b\right)^{\nu-\frac{3D}{2}}\int\prod_{j=1}^3\frac{{\rm d}^Dq_j}{i\pi^{D/2}}\,\frac{D_9^{-\nu_9}}{D_1^{\,\nu_1}D_2^{\,\nu_2}D_3^{\,\nu_3}D_4^{\,\nu_4}D_5^{\,\nu_5}D_6^{\,\nu_6}D_7^{\,\nu_7}D_8^{\,\nu_8}}\Big|_{{\rm cut}_{4b}}\,,
	\end{aligned}
\end{equation}
where $m_b$ is the bottom quark mass and $D_9$ is treated as irreducible scalar product (ISP), i.e., $\nu_9\leq 0$. The spacetime dimension $D=4-2\varepsilon$ unless otherwise specified and $\nu =\sum_i \nu_i$.
The inverse propagators are defined as
\begin{equation}
	\begin{aligned}
		D_1 &= q_1^2, \,\,\,\quad \qquad D_2 = (q_1 - q_2)^2 - m^2_b,\,\,\quad\quad \qquad D_3 = q_2^2 - m^2_b,\\
		D_4 &= q_3^2,\,\,\,\quad \qquad D_5 = (q_2 + q_3)^2 - m^2_b,\,\,\quad\quad \qquad D_6 = (q_3 - k_1)^2,\\
		D_7 &= (q_1 + k_1)^2,\,\,\, D_8 = (q_1 - q_2 - q_3 + k_1)^2 - m^2_b,\,\,\,D_9 = (q_1 - q_2 - q_3)^2 - m^2_b.
	\end{aligned}
\end{equation} 
Here $k_1$ is the external momentum satisfying the on-shell condition $k_1^2 = m_H^2$. We are interested in the contribution to the decay $H\rightarrow b\bar{b}b\bar{b}$, therefore we need to enforce cuts on the four massive propagators, i.e., $D_2, D_3, D_5$ and $D_8$, as indicated by the subscript ``${\rm cut}_{4b}$''. The integrals contain two mass scales. After normalization, the results depend on a single dimensionless variable, 
\begin{equation}
	z = \frac{m_H^2}{m^2_b} \,,
\end{equation}
which is about $\mathcal{O}(10^3)$ for physical parameters.

\begin{figure}[!htp]
	\centering
	\includegraphics[width=14cm]{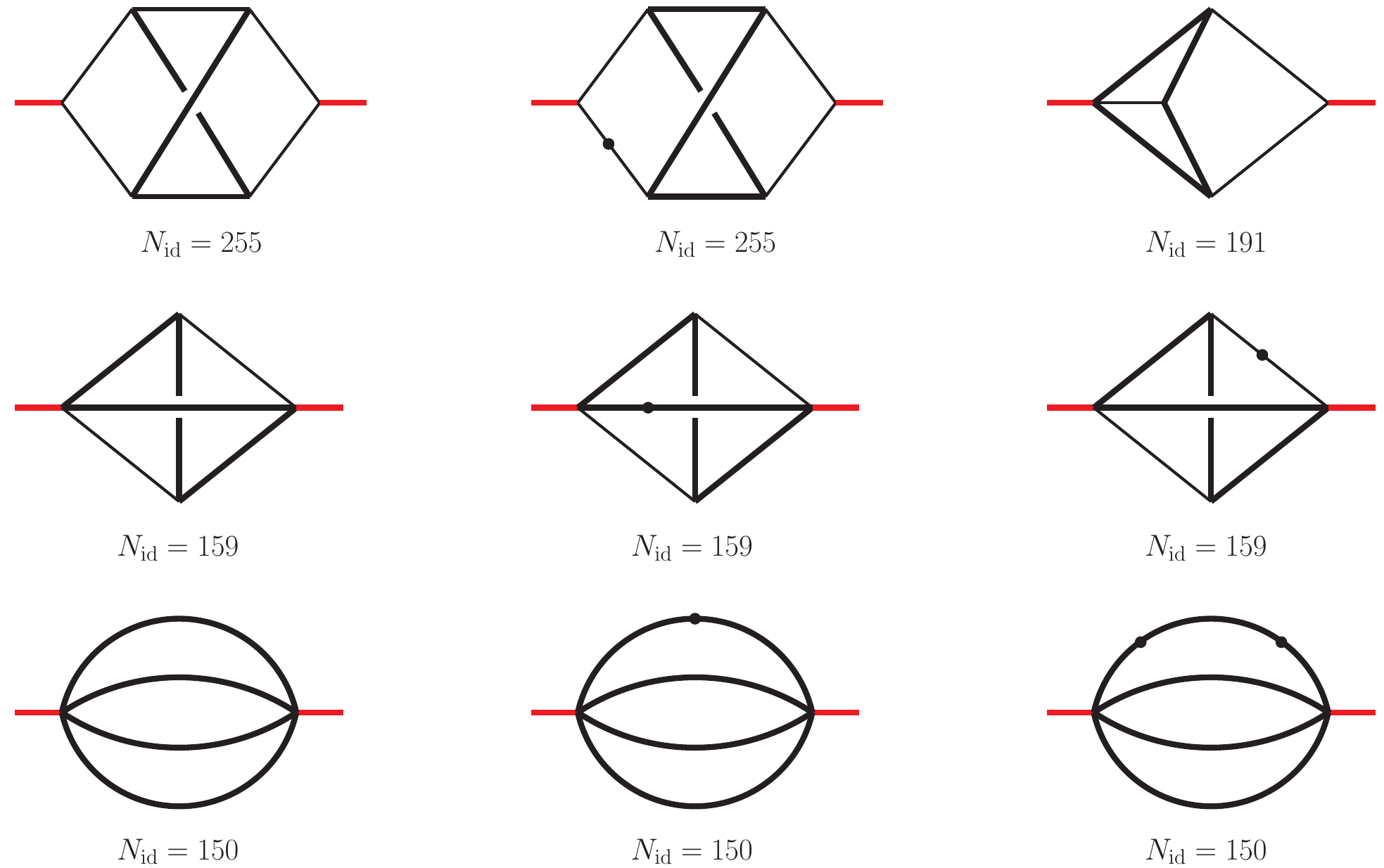}
	\caption{\label{fig:higgs3L}Four sectors in this integral family under the four massive cuts. The thick black and red lines stand for the massive
bottom quark and the Higgs boson, respectively, while the thin lines denote the gluon. One black dot indicates one additional power of the
corresponding propagator.}
\end{figure}

In total, nine master integrals survive under the four massive cuts in four sectors. The sector ID is defined as
\begin{equation}
    N_{\rm id} = \sum_{n=1}^9 \Theta(\nu_n)\,2^{n - 1}
\end{equation}
with the Heaviside step function $\Theta(x)=1$ or 0 for $x>0$ and $x\le 0$, respectively.
The four relevant sectors with sector ID $150, 159, 191, 255$ have three, three, one, and two master integrals, respectively. The topologies of these nine master integrals are shown in Fig.~\ref{fig:higgs3L}. 
The top sector, with the sector ID $N_{\rm id} = 255$, has an elliptical curve, which is the same as the one in a two-loop non-planar triangle family~\cite{Jiang:2023jmk}. The next-to-top sector, with the sector ID $N_{\rm id} = 191$, is the three-loop kite integral, which has been investigated in \cite{Wang:2024ilc}. 
This sector has only one master integral, which can be expressed in terms of MPLs.
The next-to-next-to-top sector, with the sector ID $N_{\rm id} = 159$ is also of the MPL type under its maximal cut, but is non-trivial. We have proposed a procedure to find the $\varepsilon$-factorized differential equation in such a sector based on the Picard-Fuchs operators in \cite{Wang:2024ilc}. Finally, we have the three-loop equal-mass banana sector, corresponding to $N_{\rm id} = 150$, which was analyzed in \cite{Pogel:2022yat} in detail.


Before we dive into calculations in each sector, it is essential to introduce the framework and required notation. The general strategy is ``separate and conquer'': we work out the $\varepsilon$-factorized differential equations sector by sector;  then, we deal with the mixing. In each sector, the algorithm consists of two steps. In step 1,  we inspect the structure of master integrands (integrands modulo linear relations) in the loop-by-loop Baikov representation. In a sector with ID $N_{\rm id}$, the loop-by-loop Baikov representation of a master integral reads
\begin{equation}
    M_i^{( N_{\rm id} )} =  C_{\rm Baikov}^{(N_{\rm id})} \int_{\mathcal{C}^{( N_{\rm id} )}} \frac{{\rm d}^{n} x}{(2\pi i)^{n}}\,u^{( N_{\rm id} )}(x, z)\,\hat{\phi}^{( N_{\rm id} )}(x, z)\,.
    \label{eq:baikov}
\end{equation}
Here, the superscript indicates the sector under consideration, and the corresponding maximal cut in this sector has been enforced.  We only need to choose one of the master integrals in the specified sector to obtain the minimal twist, $u^{( N_{\rm id} )}(x, z)$, and the overall factor $C_{\rm Baikov}^{(N_{\rm id})}$. The minimal twist in the loop-by-loop representation, which is the pivotal component in our analysis, takes the following form,
\begin{equation}
    u^{( N_{\rm id} )}(x, z) = \prod_{i \in I_{\text {odd }}}\left[p^{( N_{\rm id} )}_i(x)\right]^{-\frac{1}{2}+\frac{1}{2} b_i \varepsilon} \prod_{i \in I_{\text {even }}}\left[p^{( N_{\rm id} )}_i(x)\right]^{\frac{1}{2} b_i \varepsilon}\,,\quad b_i\in\mathbb{Z}\,,
\end{equation}
where $p_i$'s are polynomials in the remaining Baikov variables $x$, whose coefficients may contain $z$. 
We call the $p_i$'s with a square root exponent odd polynomials (with the index set $I_{\rm odd}$) and the others even polynomials (with the index set $I_{\rm even}$). 
They play different roles in the proposed algorithm. The rational part, $\hat{\phi}^{( N_{\rm id} ) }(x)$, differs for distinct master integrals $M_i$ in a given sector, and is made up of products of Baikov polynomials with integer powers:
\begin{equation}
    \hat{\phi}^{( N_{\rm id} ) }(x) = \frac{ q^{( N_{\rm id} ) }(x) }{\prod\limits_{i\in I_{\rm all}}\left[p_i^{( N_{\rm id} ) }(x)\right]^{\mu_i}}\,,\quad \mu_i\in\mathbb{Z}\,,
\end{equation}
with $I_{\rm all} = I_{\rm odd}\cup I_{\rm even}$. Here, $q^{(N_{\rm id})}(x)$ is an arbitrary polynomial to balance the degrees of the polynomial after homogenization; see some explicit examples later. There are an infinite number of $\hat{\phi}^{( N_{\rm id} ) }$. However, there exist linear relations among them, such that integrands in eq. (\ref{eq:baikov}) form a finite-dimensional vector space modulo those linear relations, denoted by $H_{\omega}^n$, just like the corresponding Feynman integrals. The above Baikov variables are affine. To take infinity into account, we homogenize all the above polynomials and denote their homogeneous versions by the corresponding capital symbols. We refer interested readers to \cite{Bree:2025tug} for details when necessary. If the sector information is understood clearly, we may suppress the superscript for brevity.

Step 1 also defines some kinematic pre-factors and $\varepsilon$ factors. These factors were usually obtained by experience, but can now be defined algorithmically \cite{e-collaboration:2025frv,Bree:2025tug}. After step 1, we will have a (pre-)canonical basis, denoted by $\vec{J}$. For many MPL-type integrals, this new basis already enjoys the desired $\varepsilon$-factorized differential equation. If not, the $\varepsilon$ dependence in the differential equation has a good lower-triangle block structure, and the highest order in $\varepsilon$ is always of order $\varepsilon$, not higher:
\begin{equation}\label{eq:epsLaurent}
    {\rm d} \vec{J} = {\bf B}(z) \vec{J} = \left[\sum_{k=-N_{\nu}}^1{\bf B}^{(k)}(z) \right] \vec{J},
\end{equation}
where ${\bf B}^{(1)}$ contains all the terms of order $\varepsilon$ in ${\bf B}$ and the $j$-th sub-diagonal block of ${\bf B}^{(k)}~(k\le 0)$ is given by the terms of order $\varepsilon^{N_{\nu}+k-j}$ ($N_{\nu}+k-j\le 0$) of the corresponding block of ${\bf B}$.  
The differentiation ${\rm d}$ is taken with respect to kinematic variables rather than the Baikov variables. In this paper, there is only one kinematic variable $z$. This specific pattern is compatible with the filtration structure. Then step 2 can remove unwanted terms algorithmically, which we will illustrate in the following.  


\section{Diagonal Blocks}
\label{sec:diag}
In this section, we deal with diagonal blocks in a sector-wise way. In this context, we simply neglect the superscript on the sector ID.  Throughout this integral family, the number of Baikov variables after enforcing the maximal cut in each sector is one; i.e., we have $N_V=1$ for all sectors, and the remaining  Baikov variable in sector 255 and sector 159 with the corresponding sector-wise maximal cuts is $x_1=D_9/m_b^2$.  

\subsection{Sector 255}
We take the integral $I_{111111110}$ as the representative integral in this sector. Its loop-by-loop Baikov representation under the maximal cut reads
\begin{equation}
    \begin{aligned}
        I_{111111110}\big|_{{(255)}} = C_{\rm Baikov} \int_{\mathcal{C}}\frac{{\rm d} x_1}{2\pi i}\, u(x_1) \cdot 1,
    \end{aligned}
\end{equation}
where the pre-factor $C_{\rm Baikov}$ and the (minimal) twist function $u(x_1)$ read as
\begin{equation}
    \begin{aligned}
        C_{\rm Baikov} &= 64\pi^5 z^{-1+\varepsilon}\frac{e^{3\varepsilon\gamma_E}\,\Gamma(1-\varepsilon)}{(1-2\varepsilon)\Gamma(1-2\varepsilon)^2},\\
       u(x_1) &= \Big[ x_1 \left(x_1+z\right) \left(x_1 \left(x_1+z\right)-4 z\right) \Big]^{-\frac{1}{2}-\varepsilon}. 
    \end{aligned}
\end{equation}
Clearly, the twist function defines an elliptical curve. By the above notation, we have three (affine) odd Baikov polynomials:
\begin{equation}
    p_1(x_1) = x_1\,,\quad p_2(x_1) = x_1 + z\,,\quad p_3(x_1) = x_1(x_1+z)-4z\,.
\end{equation}

From this (affine) twist function, one can define the connection $\omega = {\rm d}\ln u = \partial\ln u/\partial x_1\,\cdot {\rm d} x_1$, by which one can calculate the dimension of $H_\omega^1$:
\begin{equation}
    \dim H_\omega^1 = 3 > 2 = \dim V^1,
\end{equation}
where $V^1$ is the vector space of the master integrals. The dimension of $V^1$ can be determined by running \texttt{Kira} \cite{Klappert:2020nbg}. 

Based on $C_{\rm Baikov}$, we define 
\begin{equation}
    C_{\rm abs}  = z\, (1-2\varepsilon)\, \varepsilon^5,
\end{equation}
such that 
\begin{equation}
    C_{\rm Baikov}\cdot C_{\rm abs} =  64 \varepsilon^5 \pi^5\left[1+\varepsilon L_z+ \varepsilon^2\left(\frac{L_z^2}{2}-\frac{7\pi^2}{12}\right)+\mathcal{O}(\varepsilon^3)\right]
\end{equation}
is pure, where $L_z=\ln z$. 

To take infinity into account, we go to the projective space, i.e., rewrite everything with the homogeneous coordinate $[x_0 : x_1]$. This can be done by replacing $x_1$ by $x_1/x_0$ everywhere. We denote all quantities of the homogeneous version in capital letters. In particular, we have
\begin{equation}
    \label{eq:twist255}
    U(x_0, x_1) = P_0^{4\varepsilon}\,P_1^{-\frac{1}{2}-\varepsilon}\,P_2^{-\frac{1}{2}-\varepsilon}\,P_3^{-\frac{1}{2}-\varepsilon},
\end{equation}
which is of homogeneous degree $-2$, and the homogenized polynomials read
\begin{equation}
    P_0 = x_0,\quad P_1 = x_1,\quad P_2 = x_1 + z\,x_0,\quad P_3 = x_1 (x_1 + z\,x_0) - 4z\,x_0^2.
\end{equation}
In the language of \cite{e-collaboration:2025frv,Bree:2025tug}, $\{P_1, P_2, P_3\}$ are odd polynomials, which define the associated leading geometry; while $P_0$ is the only even polynomial, which tells us the possible place to take a non-zero residue on the geometry. After homogenization, we denote the index set of odd polynomials by $I_{\rm odd}^0$ while that of even polynomials by $I_{\rm even}^0$. In the current example, we have 
\begin{equation}
    I_{\rm odd}^0 = \{1, 2, 3\},\,\quad I_{\rm even}^0 = \{0\}\,.
\end{equation}
With the homogeneous coordinate, the integration measure is upgraded, 
\begin{equation}\label{eq:homoeta}
    {\rm d}x_1 \,\quad \rightarrow \quad \eta = x_0 {\rm d} x_1 - x_1 {\rm d} x_0\,,
\end{equation}
which is of homogeneous degree 2. 

\begin{figure}[!htp]
    \centering
    \includegraphics[width=0.4\linewidth]{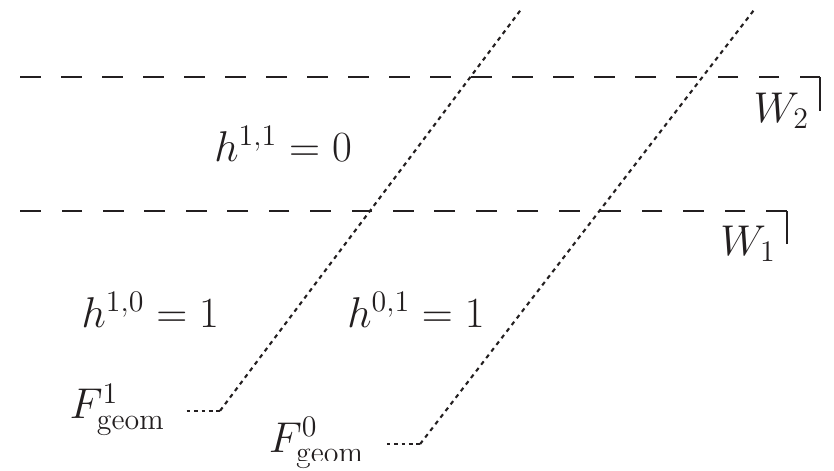}
    \caption{Organization of the vector space of master integrals in the sector 255 by two filtrations $W_w$ and $F_{\rm geom}^p$.  }
    \label{fig:255hodge}
\end{figure}

Now, we can move on to the candidates of master integrands, which constitute the linear space, denoted by $H_\omega^1$, modulo possible linear relations among them. Elements in the linear space look like
\begin{equation}
   \Psi =  C_{\rm Baikov}\cdot \underbrace{C_{\rm abs}\cdot C_{\rm rel}\cdot C_{\rm clutch}}_{C_{\varepsilon}}\,\cdot \, U \,\underbrace{\frac{Q}{P_0^{\mu_0} P_1^{\mu_1} P_2^{\mu_2} P_3^{\mu_3}}}_{\hat{\Phi}}\,\eta,
\end{equation}
where $\hat{\Phi}$ must be of homogeneous degree 0 to ensure that the homogeneous degree of $U \hat{\Phi} \eta $ is 0.  Now, we illustrate in detail how to proceed in step 1. We can have at most one residue to take since $N_V=1$. If a possible residue has been taken, the integrand is located at that point, and no more residues can be taken. 
\begin{enumerate}
    \item We start from the top layer with the weight
\begin{equation}
    w=N_V+r = 1+1=2\,,
\end{equation}
where the second ``$1$'' stands for one {\bf non-zero} residue, i.e., $r=1$. Since we have one even polynomial $P_0$, we can only take a non-zero residue at the point: $x_0=0$, i.e., $[x_0 : x_1] = [0:1]$, which corresponds to infinity in the original affine version. Hence, we have one candidate master integrand: $\hat{\Phi}=x_1/P_0$, here $Q=x_1$ to ensure that $\hat{\Phi}$ is of homogeneous degree 0 and $\mu_0=1$. By the definition of $C_{\rm rel}$ and $C_{\rm clutch}$ in \cite{e-collaboration:2025frv,Bree:2025tug}, we have
\begin{equation}
    \begin{aligned}
        C_{\rm rel} &= \prod_{i \in I_{\text {odd }}^0}\left(-\frac{1}{2}+\frac{1}{2} b_i \varepsilon\right)_{\mu_i}\;\; \prod_{i \in I_{\text {even }}^0}\left(\frac{1}{2} b_i \varepsilon\right)_{\mu_i} = 4\varepsilon\,,\\
        C_{\rm clutch} &= \varepsilon^{-|\mu|} = \varepsilon^{-1}\,,
    \end{aligned}
\end{equation}
with $(a)_n=\Gamma(a+1) / \Gamma(a+1-n) \text { being the falling factorial }$ and $|\mu|=\mu_0+\mu_1+\cdots$.   
Putting everything together, we have the following master integrand candidate:
\begin{equation}
    \Psi_3 =  C_{\rm Baikov}\cdot \underbrace{4 z (1-2\varepsilon) \varepsilon^5}_{C_{\rm abs}\cdot C_{\rm rel}\cdot C_{\rm clutch}} \cdot\, U\cdot\,\frac{x_1}{P_0}\cdot\, \eta\,.
\end{equation}
Clearly, this candidate has the pole order $o=1$ due to the simple pole at $x_0 = 0$. 
In the filtration language of \cite{e-collaboration:2025frv,Bree:2025tug}, the labels for this layer are given by
\begin{equation}
    p = N_V - o + r = 1\,,\quad q = o = 1\,.
\end{equation}
There are no other candidates at this layer since we have only one possibility of localizing to a point.

\item Next, we move one layer down where $w=N_V+r= 1+0$, i.e., we aim at differential forms with no non-zero residues. There are two cases: 1) the differential form is holomorphic; 2) the differential forms have poles without non-zero residue. We cannot use $w$ to distinguish these two cases, but they are specified by the pole order $o$. 
\begin{itemize}
    \item The first case is trivial:
\begin{equation}
    \Psi_1 = C_{\rm Baikov}\cdot \underbrace{ z (1-2\varepsilon) \varepsilon^5}_{C_{\rm abs}\cdot 1\cdot 1} \cdot\, U\,\cdot 1\, \cdot \eta\,.
\end{equation}
Here, $C_{\rm rel}$ and $C_{\rm clutch}$ are trivially 1. For this candidate, we have
\begin{equation}
    p = N_V - o + r = 1-0+0=1\,,\quad q = o = 0\,.
\end{equation}
\item Next, we require a differential form with poles but no residues. Such a form is easily obtained by increasing the power of odd polynomials by 1, e.g.,
\begin{equation}
    \Psi_2 = C_{\rm Baikov}\cdot \underbrace{ z (1-2\varepsilon) \varepsilon^5 \cdot \left(-\frac{1}{2}-\varepsilon\right) \cdot \varepsilon^{-1}}_{C_{\rm abs}\cdot C_{\rm rel}\cdot C_{\rm clutch}}\,\cdot \,U\,\cdot\,\frac{x_0 x_1}{P_3}\cdot \eta\,.
\end{equation}
The other choices of increasing the power of odd polynomials can be linearly reduced to this candidate. This candidate has
\begin{equation}
    p = N_V - o + r = 1-1+0=0\,,\quad q = o = 1\,.
\end{equation}
Note that the pole order of $\Psi$ is defined additively with respect to Baikov polynomials. In the univariate case, we define the pole order of $x^{-\alpha}\, {\rm d} x$ at $x=0$ for $\alpha>0$ to be $\lfloor\alpha\rfloor$, where $\lfloor \alpha\rfloor$ denotes the floor function. Hence, the pole order of ${\rm d} x / x$ at $x=0$ is 1, and the pole order of $x^{-3/2}{\rm d} x$ also reads $\lfloor 3/2\rfloor = 1$. 
\end{itemize}
\end{enumerate}
In general, there are linearly independent master integrand candidates in a given subspace, which is specified by $(w, o)$ or, equivalently, $(p, q)$. We need to solve the linear relations in each subspace, which is a simpler problem.  In the current context, we have exhausted all possible master integrand candidates in this sector. It is not hard to convert the above three candidates from the twisted cohomology side (i.e., the master integrand side) to the Feynman integral side, denoted as $\{\tilde{J}_8, J_8, J_9\}$, including all the pre-factors defined above:
\begin{equation}
    \begin{aligned}
        \Psi_3 \; &\rightarrow \;  \tilde{J}_8 = 4z (1-2\varepsilon) \varepsilon^5\,\cdot I_{11111111(-1)}  \, ,\\
        \Psi_2 \; &\rightarrow \; J_9 = z (1-2\varepsilon)\varepsilon^4\,\cdot I_{111111120}   \, ,\\
        \Psi_1 \; &\rightarrow \; J_8 = z (1-2\varepsilon)\varepsilon^5\,\cdot I_{111111110}  \, .
    \end{aligned}
\end{equation} 
It is easy to check that $\tilde{J}_8$ is proportional to $J_8$, which is not revealed in the above-described procedure\footnote{This can also be investigated from the twisted cohomology side.}. After the above graded procedures, we have decomposed the original vector space of master integrands, hence that of master integrals, into subspaces according to filtrations. The filtration ordering numbers $(w, o)$, or equivalently $(p, q)$, decompose the vector space into subspaces, denoted as $H_\omega^{(p,q)}$. We denote the corresponding dimension of these subspaces as $\dim H^{(p, q)}_\omega = h^{p, q}$ and organize them in the following Hodge-like triangle, shown in Fig. \ref{fig:255hodge}. 
For details of the definitions of these filtrations, please refer to \cite{Bree:2025tug}.  

The new basis $(J_8, J_9)^T$ is a good basis in sector 255, and its differential equation takes the form in \eqref{eq:epsLaurent} with the connection matrix 
\begin{equation}
    {\bf B}_{255} = \begin{bmatrix}
        {\color{red}-\frac{1}{z}} - \frac{3\varepsilon}{z} &\hspace{1cm} -\frac{4\varepsilon}{z}\\
        {\color{red}\frac{1}{\varepsilon}\frac{1}{z(z+16)}}{\color{blue}+ \frac{6}{z(z+16)}} + \frac{8\varepsilon}{z(z+16)} &\hspace{1cm} {\color{red}-\frac{1}{z+16}} -\frac{\varepsilon}{z+16}
    \end{bmatrix}{\rm d} z\,.
    \label{eq:B255full}
\end{equation}
We can read off two unwanted matrices (colored in the above equation) with different $B$-ordering~\footnote{See \cite{Bree:2025tug} for the detailed definition.} as:
\begin{equation}
    \label{eq:B255}
    {\bf B}_{255}^{(-1)} = \begin{bmatrix}
        {\color{red}-\frac{1}{z}}  &\hspace{1cm} {\color{red}0}\\
        {\color{red}\frac{1}{\varepsilon}\frac{1}{z(z+16)}} &\hspace{1cm} {\color{red}-\frac{1}{z+16}}
    \end{bmatrix}{\rm d} z,\quad {\bf B}_{255}^{(0)} = \begin{bmatrix}
        {\color{blue}0}  &\hspace{1cm} {\color{blue}0}\\
        {\color{blue} \frac{6}{z(z+16)}} &\hspace{1cm} {\color{blue}0}
    \end{bmatrix}{\rm d} z.
\end{equation}
In step 2, we introduce two rotation matrices according to the pattern of \eqref{eq:B255} to remove the above unwanted parts\footnote{The general algorithm is described in Section 4.2 of \cite{Bree:2025tug}.}:
\begin{equation}
    {\bf R}_{255}^{(-1)}{\bf R}_{255}^{(0)} \begin{pmatrix}
        K_8\\
        K_9
    \end{pmatrix} = \begin{pmatrix}
        J_8\\
        J_9
    \end{pmatrix}.
    \label{eq:KJrelation}
\end{equation}
We first perform the rotation ${\bf R}_{255}^{(-1)}$ and then ${\bf R}_{255}^{(0)}$.
Their structures are inferred according to \eqref{eq:B255} as
\begin{equation}
    {\bf R}_{255}^{(-1)} = \begin{bmatrix}
         R_{255,\, 11}^{(-1)} (z) &\hspace{0.5cm}  0\\
        \frac{1}{\varepsilon}\,R_{255,\, 21}^{(-1)} (z) &\hspace{0.5cm} R_{255,\, 22}^{(-1)}(z)
    \end{bmatrix},\quad {\bf R}_{255}^{(0)} = \begin{bmatrix}
         1 &\hspace{0.5cm}  0\\
        \,R_{255,\, 21}^{(0)} (z) &\hspace{0.5cm} 1
    \end{bmatrix}.
    \label{eq:R255b}
\end{equation}
The above matrix elements can be determined as follows.
\begin{enumerate}
    \item After performing the inverse of ${\bf R}_{255}^{(-1)}$ on $(J_8, J_9)^T$, there is an $\varepsilon^0$ term in the position $(1, 1)$ of the new connection matrix. To remove this term, we need
    \begin{equation}
        R_{255,\, 21}^{(-1)} (z) = -\frac{1}{4}\frac{\rm d}{{\rm d} z} \left(z\,R_{255,\, 11}^{(-1)} (z)\right)\,.
    \end{equation}
    To remove the $\varepsilon^{-1}$ term in the position $(2, 1)$ of the new connection matrix, we get a constraint only on $R_{255,\, 11}^{(-1)} (z)$:
    \begin{equation}\label{eq:PFE}
    \hat{L}_2(z)\,R_{255,\, 11}^{(-1)} (z) = 0\,,
\end{equation}
where $\hat{L}_2(z)$ is nothing but the Picard-Fuchs operator
\begin{equation}
    \label{eq:PF}
    \hat{L}_2(z) = \frac{{\rm d}^2}{{\rm d} z^2} + \left(\frac{2}{z}+\frac{1}{z+16}\right)\frac{{\rm d}}{{\rm d} z}+\frac{z+4}{z^2(z+16)}\,.
\end{equation}
This operator can be derived either by imposing the above requirement to remove unwanted terms in ${\bf B}_{255}^{(-1)}$ or directly by the Picard-Fuchs approach described in \cite{Wang:2024ilc}.  In this uni-kinematic variable case, they are essentially equivalent. However, for general cases with more than one kinematic variable, the Picard-Fuchs operator shall be replaced by the Picard-Fuchs ideal, but the current method still applies. For the proof-of-concept of a more complicated example, see \cite{Pogel:2025bca}. We postpone the solution of this constraint at the moment and assume that $R_{255,\, 11}^{(-1)} (z)$ is a known function to proceed. Substituting the above constraints into the connection matrix, we find that there is still a $\varepsilon^{0}$ term sitting in the $(2, 2)$ position of the new connection matrix. 
Requiring that this term vanish, we get
\begin{equation}
    R_{255,\, 22}^{(-1)}(z) = \frac{1}{z(z+16) R_{255,\, 11}^{(-1)}(z) }\,.
\end{equation}
Now, we have finished the first rotation. 
\item We then implement the second rotation in (\ref{eq:R255b}). 
Adopting the above ansatz of the basis after the first rotation, we find that the $(2, 1)$ matrix element contains a $\varepsilon^{0}$ term.
Consequently,  we have to solve a first-order inhomogeneous differential equation to remove it, which leads to
\begin{equation}
    R_{255,\, 21}^{(0)} (z) = -\frac{z(z+24)}{4} \Big( R_{255,\, 11}^{(-1)}(z)\Big)^2\,.
\end{equation}

\end{enumerate}

Summarizing the above results,  we derive from (\ref{eq:KJrelation}) the following new basis,
\begin{equation}\label{eq:K8K9}
      \begin{aligned}
      K_8 &= \frac{J_8}{R_{255,\, 11}^{(-1)}(z)},\\
          K_9 &= \left[-\frac{z(z+16)}{\varepsilon}\frac{\rm d}{{\rm d} z}\Big(z\cdot R_{255,\, 11}^{(-1)}(z)\Big) + \frac{z(z+24)}{4}R_{255,\, 11}^{(-1)}(z)\right]J_8 +z(z+16)R_{255,\, 11}^{(-1)}(z)\, J_9\,,
      \end{aligned}
\end{equation}
whose differential equation under the maximal cut is canonical, 
i.e., only the terms proportional to $\varepsilon$ survive in the connection matrix.
Note that the above basis can be related to the derivative basis \cite{Jiang:2023jmk}, which has been introduced in the calculation of the sunrise diagram. 

The only unknown matrix element is now $R_{255,\, 11}^{(-1)}(z)$, which can be derived from \eqref{eq:PFE}.
It is well known that this equation admits two independent solutions (periods). 
One can resort to the Frobenius method to solve it around a regular singularity. 
Since $z$ is hundreds for physical parameters, it is natural to solve the differential equation around $z=\infty$. 
Here, we present the results of two independent solutions \footnote{Interested readers can refer to, e.g., \cite{Jiang:2023jmk} for details of the Frobenius method and the series expansion procedure.} with the convergent radius being $|z|>16$ as follows:
\begin{equation}\label{eq:psitoptilde}
    \begin{aligned}
        \tilde{\psi}_0(z) &= -\frac{1}{z}\sum_{n\geq 1}\binom{2n}{n}^2 \left(-\frac{1}{z}\right)^{n} =\frac{1}{z}\, \frac{2}{\pi}\,K\left(k^2\right)\,,\quad\text{with}\quad k^2=-\frac{16}{z}\,,\\
        \tilde{\psi}_1(z) &= \frac{1}{2\pi i}\left[\tilde{\psi}_0\ln \left(\frac{1}{-z}\right) + \Sigma_2(z)\right] = \frac{1}{z}\, \frac{2}{\pi}\left[\frac{i}{2} K\left(1-k^2\right) - 2 K(k^2)\right]\,,
    \end{aligned}
\end{equation}
with $\Sigma_2(z)$ a holomorphic function around $z=\infty$. Here, $K(x)$ is the complete elliptic function of the first kind. The first solution is holomorphic around $z=\infty$ while the second one contains logarithmic terms. For physical parameters, $|z|\gg 16$, and thus we simply choose $R_{255,\, 11}^{(-1)}(z)$ to be holomorphic $\tilde{\psi}_0$ and proceed. With these two solutions at hand, it is useful to define a new variable:
\begin{equation}\label{eq:qtoptilde}
   \tilde{q}=e^{2\pi i \tilde{\tau}},\quad \text{with}\quad \tilde{\tau} = \frac{\tilde{\psi}_1}{\tilde{\psi}_0}\,.
\end{equation}
Substituting \eqref{eq:psitoptilde} into the above equation, we arrive at
\begin{equation}\label{eq:qtoptildeexpansion}
    \tilde{q} = -z^{-1} + 8\, z^{-2} - 84\, z^{-3} + 992\, z^{-4} - 12514\, z^{-5} + \mathcal{O}\big(z^{-6}\big)\,. 
\end{equation}
From the above equation, we could express $z$ in terms of $\tilde{q}$:
\begin{equation}
    z = -\frac{1}{\tilde{q}}-8-20\, \tilde{q} + 62\, \tilde{q}^3 - 216\, \tilde{q}^5 + \mathcal{O}\big(\tilde{q}^6\big)\,.
\end{equation}
Note that the above two series expansions have finite convergent radii, which cover the physical parameter regime.

These two series solutions of (\ref{eq:PFE}) can be analytically continued beyond the convergent radius $|z|>16$ via the property of complete elliptic functions; see, e.g., \cite{Jiang:2023jmk}. Following the same logic therein, we have two other independent solutions
\begin{equation}
\begin{pmatrix}
    \psi_1(z)\\
    \psi_0(z)
\end{pmatrix} =  \,\boldsymbol{\gamma}(z) \cdot  \begin{pmatrix}
    \tilde{\psi}_1(z)\\
    \tilde{\psi}_0(z)
\end{pmatrix}\,,
\end{equation}
and the monodromy matrix $\boldsymbol{\gamma}(z)$ is given by\footnote{Here, the point $z=-8$ is where the modulus $k^2$ equals to 2.}
\begin{equation}
    \boldsymbol{\gamma}(z)= \begin{cases}\begin{pmatrix}
        1 & 2 \\
0 & 1
    \end{pmatrix}, & z>-8 \,, \\
\begin{pmatrix}
    1 & 0 \\
0 & 1
\end{pmatrix}, &  z \leq -8 \,.
\end{cases}
\end{equation}
Accordingly, one may define the mirror variable extensible to all kinematic regions,

It was shown in \cite{Jiang:2023jmk} that a closed form of the inverse map exists, given by
\begin{equation}\label{eq:ztau}
    z(\tau) = -\frac{\eta(2 \tau)^{24}}{\eta(\tau)^8 \eta(4 \tau)^{16}} = -\frac{16}{\lambda(2 \tau)}\,,\quad\text{with}\quad \tau=\frac{\ln q}{2\pi i}\,,
\end{equation}
where $\eta(\tau)$ is the Dedekind eta function, \textit{not} the homogeneous measure $\eta$ in \eqref{eq:homoeta}, and $\lambda(\tau)$ is the modular $\lambda$ function. 
It is easy to verify that \eqref{eq:qtop} and \eqref{eq:ztau} reproduce the previous expanded results. 
Now, for any $z\in\mathbb{R}+i0^+$, it can be proven that $\operatorname{Im}(\tau)>0$ by \eqref{eq:ztau}, i.e., $\tau$ is a point on the upper-half complex plane, denoted by $\mathbb{H}$, and vice versa:
\begin{equation}
    z\in\mathbb{R}+i0^+ \,\xrightarrow{ \eqref{eq:qtop}}\, \tau(z) \in \mathbb{H} \,\xrightarrow{ \eqref{eq:ztau}}\, z\in\mathbb{R}+i0^+. 
\end{equation}
In other words, we can interchange $z$ and $\tau$ via \eqref{eq:qtop} and \eqref{eq:ztau} freely. 
Actually, the above maps can be extended to $z\in\mathbb{C}$ where the physical parameter is located on a line. 
In practice, it is more convenient 
to use $q=e^{2\pi i \tau}$ rather than $\tau$. Then, the upper half complex plane $\mathbb{H}$ is mapped to the unit disk $\mathbb{D}$. The function $z(\tau)$ is closely related to the congruence subgroup $\Gamma_0(4)$ of ${\rm SL}(2, \mathbb{Z})$:
\begin{equation}
    \Gamma_0(4) = \left\{\begin{pmatrix}
        a\; &\; b\\
        4k\; & \; d
    \end{pmatrix}\;\big|\; a, b, k \in \mathbb{Z},\,ad-4k b = 1 \right\}\,.
\end{equation}
It can be checked explicitly that $\forall\, \gamma\in\Gamma_0(4)$, $z(\tau)$ is invariant under the following transformation:
\begin{equation}
    z\big(\gamma\, \tau\big) = z(\tau)\,,\quad \text{with}\quad \gamma\, \tau = \frac{a \, \tau + b}{4k\, \tau + d}\,.
\end{equation}

\begin{equation}\label{eq:qtop}
   q(z)=e^{2\pi i \tau(z)},\quad \text{with}\quad \tau(z) = \frac{\psi_1(z)}{\psi_0(z)}.
\end{equation}
\begin{figure}[!htp]
    \centering
    \includegraphics[width=1\linewidth]{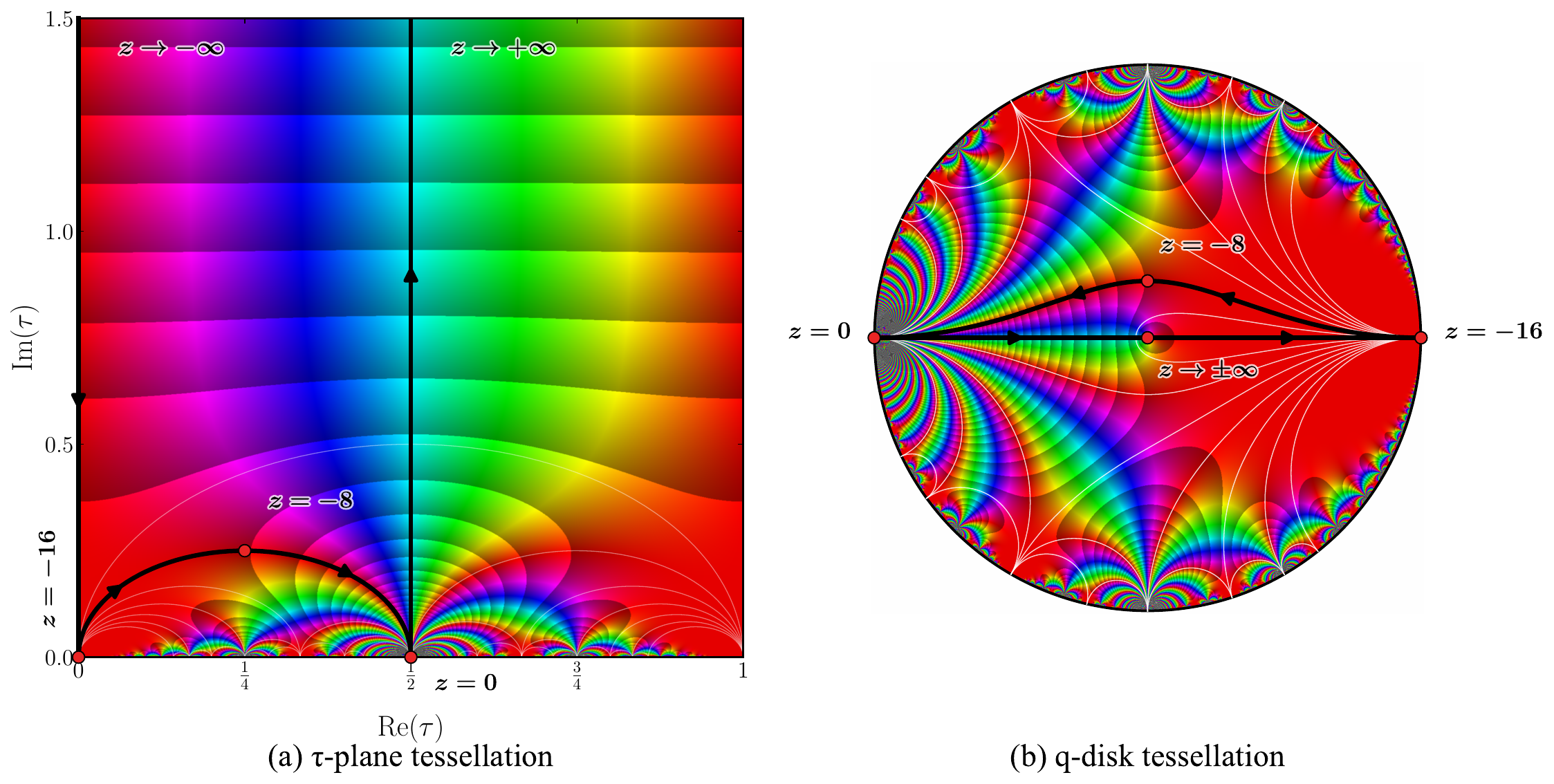}
    \caption{ Modular tessellation of $z(\tau)$ for $\tau\in\mathbb{H}$  (the left panel) and of $z(q)$ for $q\in\mathbb{D}$  (the right panel). 
    The black thick lines indicate the kinematic flow of $z\in\mathbb{R}+i0$ inside the tessellation. The hue records the complex phase $\arg(z)$ while the brightness records the magnitude $|z|$. So points with similar colors have similar complex phases and magnitudes. This plot is initially generated by \texttt{Mathematica} using \eqref{eq:ztau} as a benchmark, then passed to ChatGPT 5.5 Pro to convert into Python. The source code is uploaded with the preprint as an attachment.  }
    \label{fig:tessellation}
\end{figure}

In addition, by the property of the $\lambda$ function, $z(\tau)$ is holomorphic in $\mathbb{H}$. As a result, $z(\tau)$ is, in fact, a modular function of $\Gamma_0(4)$. The congruence subgroup $\Gamma_0(4)$ has three cusps: $\{0, 1/2, i\infty\}$, which are not in $\mathbb{H}$. These three cusps correspond to three kinematic singularities in this sector:
\begin{equation}
    \begin{aligned}
        \tau(-16+i0^+) = 0\,,\quad\tau(0+i0^+) = \frac{1}{2}\,,\quad  \tau(\infty) = i\infty\,,
    \end{aligned}
\end{equation}
by \eqref{eq:qtop}
and
\begin{equation}
    z(0) = -16 + i 0\,,\quad z(1/2)  = +i0\,,\quad z(i\infty) = \infty\,.
\end{equation}
by \eqref{eq:ztau}. Note that it is slightly different from the convention in \cite{Jiang:2023jmk}, which is equivalent up to a modular transformation. Interested readers can refer to \cite{Weinzierl:2022eaz,Broedel:2018rwm} for an introduction to modular forms and modular functions in the Feynman integral context. We visualize this modular function in Fig. \ref{fig:tessellation} to further show the correspondence and explain why the three points $\{0, 1/2, i\infty\}$ are called ``cusps''.

Now we present the exact result of the analytically continued $R_{255,\, 11}^{(-1)}(z)$, chosen as the holomorphic period and expressed in terms of $\tau$:
\begin{equation}
    R_{255,\, 11}^{(-1)}\big(z(\tau)\big) = \psi_0\big(z(\tau)\big) =  -\frac{\eta(\tau)^4\,\eta(4\tau)^{12}}{\eta(2\tau)^{14}}\,.
\end{equation}
It turns out\footnote{This can be seen by rewriting \eqref{eq:K8K9} in the derivative basis. } that $R_{255,\, 22}^{(-1)}(z)$ is related to the Jacobian $J(z)$:
\begin{equation}
   J(z) \frac{\rm d}{{\rm d} z} = \frac{\rm d}{{\rm d}\tau }\,, \quad\text{with}\quad J(z) = -\frac{z\,R_{255,\, 11}^{(-1)}(z)}{R_{255,\, 22}^{(-1)}(z)} = -z^2 (z+16)\,\psi_0(z)^2 = \frac{\eta(2\tau)^{20}}{\eta(4\tau)^{16}}\,,
\end{equation}
which is consistent with \eqref{eq:ztau}.

\subsection{Sector 191}
This sector does not have its own loop-by-loop representation\footnote{One can, however, choose another master integral that does admit a loop-by-loop Baikov representation. }, but rather inherits the loop-by-loop Baikov representation from sector 255. But in the current context, one can only take 7 existing propagators onshell instead of 8. Then, the master integral in this sector inherits the same $C_{\rm Baikov}$ and thus the chosen master integrand should be equipped with the same $C_{\rm abs}$ to make their product pure. With this pre-factor, the differential equation is already canonical, including the mixing part! This phenomenon occurs frequently in the MPL sectors. As a result, we have the third canonical master integral given by
\begin{equation}
    K_7 = z(1-2\varepsilon)\varepsilon^5 I_{111111010}. 
\end{equation}
This integral, called the three-loop kite integral, was investigated using another approach in \cite{Wang:2024ilc}. 

\subsection{Sector 159}
This sector is of the MPL type by its own and has three basis integrals. It is better here to work with $D=2-2\varepsilon$ spacetime dimensions, since this choice makes the discussion easier without losing any generality.  We use one of the master integrals in this sector to derive the (minimal) twist,
\begin{equation}
    \begin{aligned}
        I^{D=2-2\varepsilon}_{111110010}\big|_{(159)} = C_{\rm Baikov}\int_{\mathcal{C}}\frac{{\rm d} x_1}{2\pi i}\, \tilde{u}(x_1) \cdot 1,
    \end{aligned}
\end{equation}
where the pre-factor $C_{\rm Baikov}$ and the twist $\tilde{u}(x_1)$ read 
\begin{equation}
    \begin{aligned}
        C_{\rm Baikov} &= -64\varepsilon\, \pi^4 z^{\varepsilon}\frac{e^{3\varepsilon\gamma_E}\,\Gamma(1-\varepsilon)}{\Gamma(1-2\varepsilon)^2},\\
       \tilde{u}(x_1) &= x_1^{-2-2\varepsilon} (x_1 - 1)^{\varepsilon}\Big[ x_1^2 + 2 z x_1 + z(z-4) \Big]^{-\frac{1}{2}-\varepsilon}. 
    \end{aligned}
\end{equation}
Note that there is $-2$ in the exponent of $x_1$ in the twist function $\tilde{u}(x_1)$. 
Thus, $\tilde{u}^{(159)}$ is a non-minimal twist. 
One can always move this part into $\hat{\phi}$ to obtain the minimal twist. Or equivalently, we can look at the Baikov representation of $I_{11111001(-2)}\big|^{D=2-2\varepsilon}_{(159)}$. After all, the minimal twist simply reads
\begin{equation}
    u(x_1) = x_1^{-2\varepsilon} (x_1 - 1)^{\varepsilon}\Big[ x_1^2 + 2 z x_1 + z(z-4) \Big]^{-\frac{1}{2}-\varepsilon}.
\end{equation}
With this twist, we can determine the dimension of $H_\omega^1$ in this sector:
\begin{equation}\label{eq:dim159}
    \dim H_\omega^1 = 3 = \dim V^1,
\end{equation}
where $V^1$ is the vector space of master integrals in this sector \footnote{ $H_\omega^1$ here is not the same as the previous one. Neither is $V^1$. We abuse the notation a bit.}. 
The above identity indicates that there is an isomorphism between the twisted cohomology space and the Feynman integral space. 

As in the previous sectors, we upgrade everything in the homogeneous coordinate, and have
\begin{equation}
    \begin{aligned}
        U(x_0, x_1) &= P_0^{3\varepsilon}P_1^{-2\varepsilon} P_2^{\varepsilon}P_3^{-\frac{1}{2}-\varepsilon},\\
        P_0 = x_0,\quad P_1 = x_1,\quad P_2 &= x_1 - x_0,\quad P_3 = x_1^2 +2z x_1 x_0 + z(z-4)x_0^2. 
    \end{aligned}
\end{equation}
It is clear that  we have three even polynomials, which provide three points to take potentially non-zero residues:
\begin{equation}\label{eq:3local}
    P_0\,\, \leadsto\,\, [0:1],\quad P_1\,\, \leadsto\,\, [1:0],\quad P_2\,\, \leadsto\,\, [1:1],
\end{equation}
and only one odd polynomial $P_3$. 

\begin{figure}[!htp]
    \centering
    \includegraphics[width=0.4\linewidth]{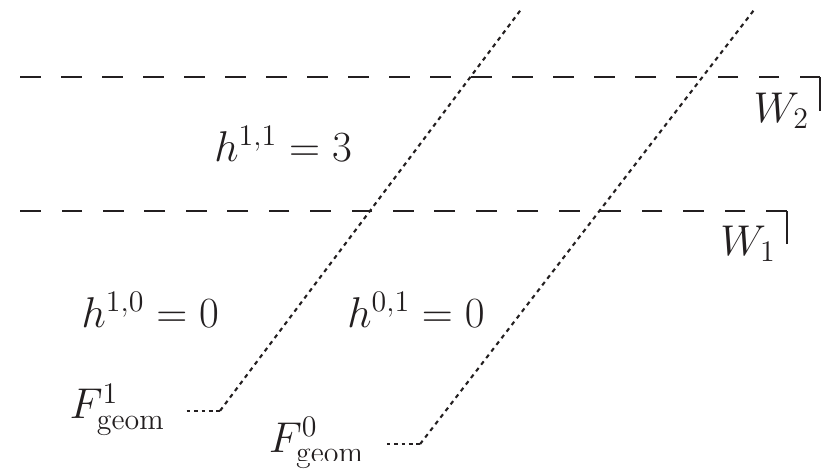}
    \caption{Organization of the vector space of master integrals in the sector 159 by two filtrations $W_w$ and $F_{\rm geom}^p$.   }
    \label{fig:159hodge}
\end{figure}

Based on $C_{\rm Baikov}$ given above, we can read off $C_{\rm abs}$ for this sector:
\begin{equation}
    C_{\rm abs} = \varepsilon^3,
\end{equation}
such that their combination is pure. The fact that $\deg\, U = -1$ and $\deg\, \eta =2$ requires that $\deg \hat{\Phi} = -1$. As in the previous discussion, we start from the top layer with the Hodge weight
\begin{equation}
    w=N_V+r = 1+1=2\,.
\end{equation}
Since we have three possible points to take non-zero residues, as shown by \eqref{eq:3local}, we construct the following three candidates:
\begin{equation}\label{eq:MPsis159}
    \begin{aligned}
         \Psi_4 &= C_{\rm Baikov}\cdot \varepsilon^3 \cdot 3\varepsilon \cdot \varepsilon^{-1}\cdot U\cdot\frac{1}{P_0}\cdot \eta\, ,\\
        \Psi_5 &= C_{\rm Baikov} \cdot \varepsilon^3 \cdot (-2\varepsilon) \cdot \varepsilon^{-1}\cdot U \cdot\frac{1}{P_1}\cdot \eta\, ,\\
        \Psi_6 &= C_{\rm Baikov} \cdot \varepsilon^3 \cdot \varepsilon \cdot \varepsilon^{-1}\cdot U \cdot\frac{1}{P_2}\cdot \eta \,.
    \end{aligned}
\end{equation}
There are no linear relations among these three candidates at this layer. Then we move one layer down to $w=N_V+r= 1+0$, i.e., where we search for differential forms with no non-zero residues. Since there are no holomorphic candidates of degree $-1$, we only need to consider the singular candidates with zero residues, which can be found by raising the exponent of the only odd polynomial, e.g.,
\begin{equation}
    \Psi'_6 = C_{\rm Baikov}\cdot \varepsilon^3 \cdot \left(-\frac{1}{2}-\varepsilon\right)\cdot\varepsilon^{-1}  \cdot U \cdot\frac{z\,x_0 - x_1}{P_3}\cdot  \eta\,,
\end{equation}
where $q=z\,x_0 - x_1$ is chosen to maintain the homogeneous degree. However, this integrand is linearly expressible in terms of the above basis  \eqref{eq:MPsis159} as:
\begin{equation}
    \Psi'_6 = \frac{(z-1)\big(2+\varepsilon(z+8)\big)}{\varepsilon\,z(z-1)^2} \Psi_5 - \frac{3(z+2)}{2z(z-1)^2} \Psi_6\,. 
\end{equation}
This relation can be easily derived both in the Feynman integral and in the twisted cohomology spaces. Therefore, no independent candidates are found at this layer. This is in line with the fact that $\dim V^1 = \dim H_\omega^1=3$ in this sector. In general, we should repeat what we have done for the top sector. 
In summary, we have the following Hodge-like triangle shown in Fig. \ref{fig:159hodge} to represent the dimensions of decomposed subspaces in this sector. 

Translating back to the Feynman integral space, we find that
\begin{equation}
    \begin{aligned}
        \Psi_4 \; &\rightarrow \;  J_4 = 3 \varepsilon^3\,\cdot I_{11111001(-2)}^{D=2-2\varepsilon}  \,,\\
        \Psi_5 \; &\rightarrow \; J_5 = -2 \varepsilon^3\,\cdot I_{11111001(-1)}^{D=2-2\varepsilon}   \,,\\
        \Psi_6 \; &\rightarrow \;   J_6 = \varepsilon^4(1-2\varepsilon)\,z\,  I_{111110020}  \, .
    \end{aligned}
\end{equation}
The first two integrals can be transformed to the basis in $D=4-2\varepsilon$ straightforwardly.

In this new basis, the connection matrix in the diagonal sector reads
\begin{equation}
    {\bf B}_{159} = \begin{bmatrix}
        {\color{blue}\frac{2-z}{(z-4) z}}-\frac{4 (z-1) \epsilon }{(z-4) z} &\hspace{0.5cm} 0 &\hspace{0.5cm} -\frac{12 \epsilon }{(z-4) z} \\
 \frac{3 \epsilon }{2} &\hspace{0.5cm} -\frac{\epsilon }{z} &\hspace{0.5cm} \frac{6 \epsilon }{z} \\
 \frac{\epsilon }{2} &\hspace{0.5cm} 0 & \hspace{0.5cm}\frac{\epsilon }{z}
    \end{bmatrix}{\rm d} z\,.
\end{equation}
In this sector, after step 1, we only have $B$-order 0 unwanted terms: 
\begin{equation}
    \label{eq:B159}
    {\bf B}_{159}^{(0)} = \begin{bmatrix}
        {\color{blue}\frac{2-z}{(z-4) z}}  &\hspace{0.5cm} {\color{blue}0}  &\hspace{0.5cm} {\color{blue}0} \\
        {\color{blue}0}  &\hspace{0.5cm}{\color{blue}0}  &\hspace{0.5cm} {\color{blue}0}\\
        {\color{blue}0}  &\hspace{0.5cm}{\color{blue}0}  &\hspace{0.5cm} {\color{blue}0}
    \end{bmatrix}{\rm d} z\,.
\end{equation}
The structures of these matrices dictate that $J_5$ and $J_6$ are already the canonical basis, while $J_4$ needs a rescaling.
It is rather trivial to find that
\begin{equation}
    K_4 = \sqrt{z(z-4)}\, J_4,\qquad K_5 = J_5,\qquad K_6 = J_6\,,
\end{equation}
are the sought-after basis integrals in this sector. This sector, on its own, is of the MPL type. Its differential equations involve a square root, which can be rationalized. 

\subsection{Sector 150 (banana)}
We do not repeat the calculation to find the canonical basis in this sector. The algorithm will lead us to the same result in \cite{Pogel:2022yat}. The differential equation for the banana sector under the four massive cuts (the maximal cut for banana integrals) is the same as the uncut banana integrals, modulo the tadpole. As such, the $\varepsilon$-factorized differential equation in the current context is the $3\times 3$ block of that in \cite{Pogel:2022yat}. 
\begin{equation}\label{eq:epsbanana}
    \begin{aligned}
        \frac{1}{2\pi i}\frac{{\rm d}}{{\rm d}\tau_B}\begin{pmatrix}
            K_1\\
            K_2\\
            K_3
        \end{pmatrix} 
        = \,\varepsilon\,\begin{pmatrix}
        -f_{2, a}-f_{2, b} & 1 & 0 \\
        f_{4, b} & -f_{2, a}+2 f_{2, b} & 1 \\
        f_6 & f_{4, b} & -f_{2, a}-f_{2, b}
        \end{pmatrix}
        \begin{pmatrix}
            K_1\\
            K_2\\
            K_3
        \end{pmatrix}
    \end{aligned}\,,
\end{equation}
where $(K_1, K_2, K_3)^T$ is the derivative basis in \cite{Pogel:2022vat}, which can be obtained by the algorithm, see \cite{Pogel:2025bca} for the unequal-mass example to illustrate this. The rotation between $(K_1, K_2, K_3)^T$ and $(I_{011010010}, I_{021010010}, I_{031010010})^T$ is included in the auxiliary file. We do not show them all explicitly here, but only the first one:
\begin{equation}\label{eq:K1}
    K_1 = \varepsilon^3\frac{I_{011010010}^{D=2-2\varepsilon}}{\psi_0^B(z)}\,,
\end{equation}
where $I_{011010010}$ here is in $D=2-2\varepsilon$.  $\psi_0^B(z)$ is the related holomorphic period. In the language of the algorithm, it is nothing but the matrix element sitting in the $(1, 1)$ position of the first rotation matrix in step 2, $R^{(-2)}_{150,\, 11}$. The constraint on $R^{(-2)}_{150,\, 11}$, namely the related Picard-Fuchs operator, has another solution, denoted as $\psi_1^B(z)$. With these two periods, the variable $\tau^B$ is defined as
\begin{equation}
    \tau_B = \frac{\psi_1^B(z)}{\psi_0^B(z)}\,,
\end{equation}
and the corresponding mirror map reads
\begin{equation}
    q_B = e^{2\pi i \tau_B} \,.
\end{equation}
Here, $\psi_0^B(z)$ is the holomorphic period of the related K3 surface and $\psi_1^B(z)$ is the single-logarithmic one. They can be solved by the Frobenius method around a regular singularity. As a function of $z$, $\psi_0^B(z)$ and $\psi_1^B(z)$ can be continued analytically , which was achieved in \cite{Pogel:2022yat}. After analytic continuation, the functional form of $\tau_B$ reads: 
\begin{equation}\label{eq:qBz}
    \begin{aligned}
        \tau_B(z) = \begin{cases}
            \frac{i K\big(1-k^2_B(z) \big)}{K \big( k^2_B(z) \big)}\,,\qquad\qquad\qquad z\in (-\infty, 0] \cup [16+8\sqrt{3}\,, +\infty)\,;\\
            \frac{i K\big(1-k^2_B(z) \big)}{K \big(k^2_B(z) \big) + 2i\, K \big(1-k^2_B(z) \big)}\,,\,\,\,\qquad  z \in ( 0, 16+8\sqrt{3} )\,.
        \end{cases}
    \end{aligned}
\end{equation}
where 
\begin{equation}
    k^2_B(z) = \frac{16 t^2}{(3-t)(1+t)^3}\,,\quad\text{with}\quad t = \frac{\sqrt{16-z} - \sqrt{4-z}}{2}\,.
\end{equation}
The critical value of $z=16+8\sqrt{3}$ is the point where $k^2_B(z)=1$.

The inverse map of \eqref{eq:qBz} reads
\begin{equation}\label{eq:zqB}
    z = -\left[\frac{\eta(\tau_B)\eta(3\tau_B)}{\eta(2\tau_B)\eta(6\tau_B)}\right]^6\,.
\end{equation}
For $z\in\mathbb{R}+i0^+$, one can check explicitly that $\operatorname{Im}(\tau_B)>0$, namely $\tau_B\in\mathbb{H}$. Moreover, the following chain of maps holds due to the analytic continuation: 
\begin{equation}
    z\in\mathbb{R}+i0^+ \,\xrightarrow{ \eqref{eq:qBz}}\, \tau_B(z) \in \mathbb{H} \,\xrightarrow{ \eqref{eq:zqB}}\, z\in\mathbb{R}+i0^+\,.
\end{equation}
As in the previous elliptic sector, this allows us to freely use $z$ or $\tau_B$ in the final result. 

The matrix elements $f_{i}$'s are functions of $q_B$ and hence functions of $z$. They are meromorphic modular forms because they have poles. However, at leading order of $\varepsilon$, which does contribute to the decay width, the meromorphicity does not pose any difficulty. In fact, based on \cite{Pogel:2022yat} together with the boundary conditions given in the appendix, we find rather compact results after taking the cuts: 
\begin{equation}\label{eq:bananaK}
    \begin{aligned}
        K_1 &= \varepsilon^3 \cdot \left[\pi\cdot \ln(-q_B)^2 - \frac{\pi^3}{3}\right]  + \mathcal{O}(\varepsilon^4)\,\\
        K_2 &= \varepsilon^2 \cdot \left[2\pi\cdot \ln(-q_B) \right] + \mathcal{O}(\varepsilon^3)\,\\
        K_3 &= \varepsilon\cdot 2\pi + \mathcal{O}(\varepsilon^2)\,.
    \end{aligned}
\end{equation}
There are two comments in order. Firstly, the above leading results in $\varepsilon$ are exact without any truncations in the kinematic variable. Given any physical value of $z$, i.e., $z>16$ to support the four massive cuts, one can derive the related $q_B$ with the help of \eqref{eq:qBz}, and plug it into the above compact expressions to obtain the exact results. One can rewrite the $\varepsilon$-factorized differential equation \eqref{eq:epsbanana} in terms of $z$ instead of $\tau_B$ or $q_B$. However, it is not so straightforward to arrive at the above compact results in terms of $z$. Secondly, with the $\varepsilon$-factorized differential equations at hand and boundary conditions given later, we can derive the results to any order of $\varepsilon$. We give results at the next order of $\varepsilon$ in the auxiliary file. 

Since we need to switch back and forth between $q_B$ and $z$, it is helpful to quote the Jacobian between them:
\begin{equation}
    J_B = \frac{{\rm d} z}{{\rm d} \tau_B} = \frac{\psi_0^B}{W_B(z)}\,,
\end{equation}
where the holomorphic period, $\psi_0^B$, can also be expressed as the eta quotient 
\begin{equation}\label{eq:psi0K3}
    \psi_0^B(z) = 12\frac{\eta(2\tau_B)^4\,\eta(6\tau_B)^4}{\eta(\tau_B)^2\,\eta(3\tau_B)^2} 
\end{equation}
and $W_B(z)$ is the Wronksian
\begin{equation}
    W_B(z) = \frac{6 i}{\pi}\frac{1}{z \sqrt{4-z}\cdot\sqrt{16 -z}}\,.
\end{equation}

\section{Mixing}
\label{sec:mixing}
We have obtained canonical bases in individual sectors. 
In this section, we show that it is straightforward  to deal with mixing between different sectors, which was suppressed in \cite{e-collaboration:2025frv,Bree:2025tug}.   
Denote the basis, after the above steps 1 and 2, by $\vec{K}=(K_1, \cdots, K_9)^T$. There are only two $\varepsilon^0$ terms in the differential equation with respect to $z$: 
\begin{equation}\label{eq:mixing}
    \begin{aligned}
    {\rm d}\,K_4 &= \frac{{\rm d}\,\big((8+z)\psi_0^B (z)\big)}{6} K_1 + \varepsilon\cdot (\cdots)\,,\\
    {\rm d}\,K_5 &= -\frac{2}{3}\sqrt{\frac{z}{z-4}}\,{\rm d} \big((z+2)\psi_0^B(z)\big)\,K_1 + \varepsilon \cdot (\cdots),
    \end{aligned}
\end{equation}
where $\psi_0^B(z)$ is the holomorphic period given by \eqref{eq:psi0K3}. 
The mixing between $K_4$ and $K_1$ can then be easily eliminated by the following refined ansatz:
\begin{equation}\label{eq:K4K1mixing}
    K_4 \to K_4 + R_{41}(z)\,K_1
\end{equation}
with the following constraint and solution\footnote{We can freely choose the integration constants to make life easier.}:
\begin{equation}
       {\rm d} R_{41}(z) = \frac{{\rm d}\,\big((8+z)\psi_0^B (z)\big)}{6}
    \,\Longrightarrow \, R_{41}(z) = \frac{(8+z)\psi_0^B (z)}{6}\,.
\end{equation}
Similarly, for the mixing between $K_5$ and $K_1$, we assume the transformation
\begin{equation}\label{eq:K5K1mixing}
 K_5 \to K_5 + R_{51}(z)\,K_1\,,
\end{equation}
with $R_{51}$ satisfying
\begin{equation}
       {\rm d} R_{51}(z) = -\frac{2}{3}\sqrt{\frac{z}{z-4}}\,{\rm d} \big((z+2)\psi_0^B(z)\big)\,,
\end{equation}
to cancel the unwanted term in the second line of \eqref{eq:mixing}. 
The solution is not so obvious as $R_{41}$, since the right-hand side is not a total derivative any longer. 
In the kinematic region where $|z|$ is very large, it admits a good series solution to any desired order:
\begin{equation}
    R_{51}(z) = 48\left[\frac{1}{z} + \frac{7}{z^2} + \frac{62}{z^3} +\frac{639}{z^4} + \frac{7266}{z^5} + \frac{88138}{z^6} +  \frac{1117572}{z^7}\right] + \mathcal{O}(z^{-8})\,,
\end{equation}
which admits a hypergeometric function ${}_4F_3$ representation. 

\section{$\varepsilon$-factorized Differential Equation}
\label{sec:de}
In this section, we summarize the sought-after $\varepsilon$-factorized differential equation of the whole integral family. With some abuse of notation, we stick to using $K_4$ and $K_5$ after considering mixing in \eqref{eq:K4K1mixing} and \eqref{eq:K5K1mixing}. 
The differential equation of $\vec{K}$ is organized in the following form,
\begin{equation}\label{eq:epsDE}
    {\rm d} \vec{K} = \varepsilon\, \boldsymbol{A}(z)   \vec{K}\,,
\end{equation}
with the connection matrix $\boldsymbol{A}(z)$ independent on $\varepsilon$, 
\begin{equation}
    \boldsymbol{A}(z) = \begin{bmatrix}
        \boldsymbol{\omega}_{3\times 3}^{(150)}\;\; & \;\; \mathbf{0} \;\; & \;\; \mathbf{0} \;\; & \;\; \mathbf{0} \\[0.5em]
         \boldsymbol{\omega}_{3\times 3}^{(159, 150)} \;\; & \;\; \boldsymbol{\omega}_{3\times 3}^{(159)} \;\; & \;\; \mathbf{0} \;\; & \;\; \mathbf{0} \\[0.5em]
        \mathbf{0} \;\; & \;\; \boldsymbol{\omega}_{1\times 3}^{(191, 159)} \;\; & \;\; \boldsymbol{\omega}_{1\times 1}^{(191)} \;\; & \;\; \mathbf{0} \\[0.5em]
        \boldsymbol{\omega}_{2\times 3}^{(255, 150)}\;\; & \;\; \boldsymbol{\omega}_{2\times 3}^{(255, 159)} \;\; &  \;\; \mathbf{0}  & \;\; \boldsymbol{\omega}_{2 \times 2}^{(255)}
    \end{bmatrix}\,.
\end{equation}
All the matrix elements are one forms.  
The diagonal blocks read as follows.
\begin{equation}\label{eq:diablocks}
    \begin{aligned}
        \boldsymbol{\omega}_{3\times 3}^{(150)} &=  \begin{bmatrix}
        -f_{2, a}-f_{2, b} & 1 & 0 \\
        f_{4, b} & -f_{2, a}+2 f_{2, b} & 1 \\
        f_6 & f_{4, b} & -f_{2, a}-f_{2, b}
        \end{bmatrix}
    2\pi i\, {\rm d}\tau_B\,,\\
    \boldsymbol{\omega}_{3\times 3}^{(159)} &=  \begin{bmatrix}
        -\frac{1}{z}\;\;\; & \frac{1}{2\sqrt{z(z-4)}} \;\;\; & \frac{2}{z} \\[1em]
        0\;\;\; & -\frac{1}{z}-\frac{3}{z-4} \;\;\; & -\frac{12}{\sqrt{z(z-4)}}\\[1em]
        0\;\;\; &  \frac{1}{2\sqrt{z(z-4)}} \;\;\; & \frac{1}{z}
        \end{bmatrix}
    {\rm d} z\,,\quad \boldsymbol{\omega}_{1\times 1}^{(191)} = -\frac{{\rm d} z}{z}\,,\\
    \boldsymbol{\omega}_{2\times 2}^{(255)} &=  \begin{bmatrix}
        -\frac{2 (z+12)}{z (z+16)}\;\;\; & -\frac{4}{z^2 (z+16) \psi_0(z)^2}  \\[1em]
        -\frac{(z+8)^2 \psi_0(z)^2}{4 (z+16)}\;\;\; & -\frac{2 (z+12)}{z (z+16)} 
        \end{bmatrix}
    {\rm d} z \\
    &= \begin{bmatrix}
      \;2z(\tau)\big(z(\tau)+12\big)\psi_0\big(z(\tau)\big)^2\;\;\;  &\;\;\; 4\\[1em]
      \;\frac{1}{4}z(\tau)^2\big(z(\tau)+8\big)^2\psi_0\big(z(\tau)\big)^4\;\;\; & \;\;\; 2z(\tau)\big(z(\tau)+12\big)\psi_0\big(z(\tau)\big)^2\;
    \end{bmatrix} 2\pi i\,{\rm d} \tau\,.
    \end{aligned}
\end{equation}
The non-diagonal blocks read
\begin{equation}\label{eq:ndiablocks}
    \begin{aligned}
        \boldsymbol{\omega}_{1\times 3}^{(191, 159)} &=  \begin{bmatrix}
        \;0\quad  & \quad  0  \quad & \quad  -\frac{{\rm d} z}{z}\;
        \end{bmatrix}
    \,,\\
    \boldsymbol{\omega}_{2\times 3}^{(255, 150)} &= \begin{bmatrix}
        \;0\quad  & \quad  0  \quad & \quad  0 \;\\
        \;  \bigg(\frac{(3z+8) R_{51}}{4\sqrt{z(z-4)}} + \frac{(19z+32)\psi_0^B}{24}\bigg)\psi_0 \quad  & \quad  0  \quad & \quad  0 \;
        \end{bmatrix} {\rm d} z\,,\\
    \boldsymbol{\omega}_{2\times 3}^{(255, 159)} &= \begin{bmatrix}
        \;0\quad  & \quad  0  \quad & \quad  0 \;\\
        \;  \psi_0 \quad  & \quad   \frac{(3z+8) }{4\sqrt{z(z-4)}}\psi_0  \quad & \quad  5\psi_0 \;
        \end{bmatrix} {\rm d} z\,.      
    \end{aligned}
\end{equation}
$\boldsymbol{\omega}_{3\times 3}^{(159, 150)}$ is much lengthy and we do not report it here. It can be found in the auxiliary file. One can further perform a constant rotation in sector 159, such that its diagonal block exhibits the self-duality, first discovered in \cite{Pogel:2022ken,Pogel:2022vat} and then further investigated in \cite{Pogel:2024sdi,Duhr:2024xsy},  as other diagonal blocks. It can be verified explicitly that the above connection matrix has at most simple poles.

\section{Boundary Conditions}
\label{sec:bd}
Now the basis consisting of $K_{i},i=1,\cdots,9$, satisfies the $\varepsilon$-factorized differential equations. We need boundary conditions to solve these equations. 
We choose the point $z=\infty$, which is a regular singular point of the differential equation. Around this point, the solutions develop the maximally logarithmic pattern (the so-called MUM point) for both the banana sector and the elliptic sector. These logarithmic terms are dictated by the differential equation. Then, the full results are \textbf{indefinite integration} over $z$ with some constants. By boundary conditions, we mean those constants. Such a procedure of only keeping the remaining constants is called the trailing zero procedure.

We do not need to repeat the process to calculate boundary values in the banana sector from scratch. Here, we just recycle the full results in \cite{Pogel:2022yat} and take the discontinuity. For that purpose, we study the asymptotic result for the \textit{uncut} $I_{011010010}$ in $D=2-2\varepsilon$, which is one of the master integrals in the banana sector:
\begin{equation}\label{eq:I1}
    I_{011010010}^{D=2-2\varepsilon}\Big|^{\rm uncut}_{z\to \infty} \sim \sum_{n=0}^3 a_n\, z^{-1-n\varepsilon}\,.
\end{equation}
The $\varepsilon$-factorized basis needs to be normalized by the holomorphic period $\psi_0^B(z)$ given by \eqref{eq:psi0K3}. In this context, the asymptotic behavior of 
$K_1$, defined by \eqref{eq:K1} but without enforcing the four massive cuts, is
    \begin{align}
        K_1\big|^{\rm uncut}_{z\to \infty} &= \frac{\varepsilon^3\,e^{3\varepsilon\gamma_E}}{12}\,\left[  \frac{\Gamma(-\varepsilon)^4 \Gamma(1+3\varepsilon) }{\Gamma(-4\varepsilon)} \,(-z)^{-3\varepsilon} + 4 \frac{\Gamma(-\varepsilon)^3 \Gamma(\varepsilon)\Gamma(1+2\varepsilon) }{\Gamma(-3\varepsilon)}\,(-z)^{-2\varepsilon}  \right.\nonumber\\
    &\qquad\qquad \left.  + 6\frac{\Gamma(-\varepsilon)^2  \Gamma(\varepsilon)^2 \Gamma(1+\varepsilon) }{\Gamma(-2\varepsilon)}\,(-z)^{-\varepsilon}  + 4 \Gamma(\varepsilon)^3\, \right] + \mathcal{O}(z^{-1}),
    \label{eq:K1asy}
    \end{align} 
which has a combinatoric pattern and can be generalized to banana integrals at any loop, even with different mass configurations for the internal massive lines. Clearly, only the first three terms can develop discontinuity for $z>16$. In fact, the last term is related to the tadpole in the current normalization by $(m^2)^{\nu-3D/2}$, and the tadpole does not generate discontinuity.  Now, we can use the property
\begin{equation}
    \operatorname{Im}\left[(-z-i 0)^{-n \epsilon}\right]=-\sin (\pi n \epsilon)\, \theta(z) \, z^{-n \epsilon}
\end{equation}
to extract the imaginary part of the first three terms in \eqref{eq:K1BD}, leading us to\footnote{Here, $K_1$ is the imaginary part of $\varepsilon^3 I^{(2-2\varepsilon)}_{011010010}$.}:
\begin{align}\label{eq:K1BD}
       K_{1}\big|_{z\to \infty} &= \pi\frac{\varepsilon^3\,e^{3\varepsilon\gamma_E}}{12}\,\left[  \frac{\Gamma(-\varepsilon)^4  }{\Gamma(-4\varepsilon)\Gamma(-3\varepsilon)} \,z^{-3\varepsilon} + 4 \frac{\Gamma(-\varepsilon)^3 \Gamma(\varepsilon) }{\Gamma(-3\varepsilon)\Gamma(-2\varepsilon)}\,z^{-2\varepsilon}   + 6\frac{\Gamma(-\varepsilon)^2  \Gamma(\varepsilon)^2  }{\Gamma(-2\varepsilon)\Gamma(-\varepsilon)}\,z^{-\varepsilon}  \, \right] \theta(z)\nonumber \\
       &= \varepsilon^3 \left[L_z^2 - \frac{\pi^2}{3}\right]\pi + \varepsilon^4\left[-2L_z^3 + \frac{7\pi^2}{3}L_z - 10\zeta_3\right]\pi +\mathcal{O}(\varepsilon^5)\,,
\end{align}
with $L_z=\ln z$. All such logarithms will be generated by the $\varepsilon$-factorized differential equation. So, as the trailing-zero procedure does, the boundary constant for $K_1$ is nothing but
\begin{equation}\label{eq:K1BDconst}
    \begin{aligned}
       K_{1, \text{BD}} = \varepsilon^3 \left( - \frac{\pi^3}{3}\right) + \varepsilon^4\big( - 10\pi\zeta_3\big)+\mathcal{O}(\varepsilon^5),
    \end{aligned}
\end{equation}
and \eqref{eq:K1BD} can be viewed as a consistency condition to check whether the differential equation can generate these logarithms in the end. 
The above boundary value is enough for the banana sector. 
The boundary values of all others can easily be obtained by dimension shift relations and taking derivatives with respect to $z$. These two operations commute with taking imaginary parts.

Master integrals beyond the banana sectors are less symmetric than the banana ones. 
The easiest way to figure them out is to leave the boundary constants (after the trailing-zero procedure), and to run \texttt{AMFlow} \cite{Liu:2022chg} at a regular point, say $z=100$,  to high precision.
The comparison between the analytical expressions with some undetermined constants and the numerical values fixes those constants. 
Then, we run \texttt{AMFlow} at another value of $z$ to cross-check the results. We report the boundary constants using this technique as follows. 
\begin{equation}
    \begin{aligned}
       K_{4,\,{\rm BD}} &= \varepsilon^3\cdot 0 + \varepsilon^4\cdot \big(-4\pi\zeta_3 \big)\, +\mathcal{O}\big(\varepsilon^5\big)\,,\\
       K_{5,\,{\rm BD}} &= \varepsilon^3 \cdot \left(-\frac{4\pi^3}{3}\right) + \varepsilon^4  \cdot  \big(-8\pi\zeta_3 \big)\,+\mathcal{O}\big(\varepsilon^5\big)\,,\\
       K_{6,\,{\rm BD}} &= \varepsilon^4 \cdot \big(-12\pi\zeta_3\big) + \varepsilon^5  \cdot \left(\frac{137\pi^5}{180}\right)\,+\mathcal{O}\big(\varepsilon^6\big)\,,\\
       K_{7,\,{\rm BD}} &= \varepsilon^5  \cdot  \left(\frac{13\pi^5}{90}\right) + \varepsilon^6  \cdot  \left( \frac{44\pi^3\zeta_3}{3} + 16\pi\zeta_5\right)\,+\mathcal{O}\big(\varepsilon^7\big)\,,\\
       K_{8,\,{\rm BD}} &= \varepsilon^5  \cdot  \left(\frac{181\pi^5}{90}\right) + \varepsilon^6   \cdot   \left(\frac{262\pi^3\zeta_3}{3} + 132\pi\zeta_5\right)\,+\mathcal{O}\big(\varepsilon^7\big)\,,\\
       K_{9,\,{\rm BD}} &= \varepsilon^4  \cdot  \big(-12\pi\zeta_3\big) + \varepsilon^5  \cdot  \left(\frac{317\pi^5}{360}\right)\,+\mathcal{O}\big(\varepsilon^6\big)\,.
    \end{aligned}
\end{equation}

\section{Solutions of Master Integrals} 
\label{sec:sol}
From now on, we will write master integrals expanded in terms of $\varepsilon$:
\begin{equation}
    K_i(z) = \sum_{n} \varepsilon^n K_i^{(n)}(z).
\end{equation}
With the $\varepsilon$-factorized differential equation and the boundary constants at hand, it is rather straightforward to obtain $K_i^{(n)}$ for any $n$. In the physical region $z\sim 0.8\cdot 10^3$, we can simply expand all the letters in terms of $z$ to whatever orders and then perform the iterated integral term by term. The time consumption is negligible. 

Here, we report results in their leading order of $\varepsilon$ while results in the next-to-leading order of $\varepsilon$ are collected in the auxiliary file. $\big(K_1^{(3)}, K_2^{(2)}, K_3^{(1)}\big)^T$ are exact in terms of $q_B$ which is related to $z$ via \eqref{eq:qBz}. 
\begin{equation}\label{eq:bananaKleading}
    \begin{aligned}
        K_1^{(3)} &= \pi\cdot \ln(-q_B)^2 - \frac{\pi^3}{3} \,,\\
        K_2^{(2)} &= 2\pi\cdot \ln(-q_B) \,,\\
        K_3^{(1)} &= 2\pi \,.
    \end{aligned}
\end{equation}
The transition from $z$ to $q_B$ is also instant in time and does not pose any ambiguity. Of course, one can expand $q_B$ in terms of $z$ around infinity.  $K_4$ starts from $\varepsilon^4$:
\begin{equation}\label{eq:K4}
    \begin{aligned}
        K_4^{(3)} = 0 \,, 
    \end{aligned}
\end{equation}
while $K_5^{(3)}$ is non-trivial. 
\begin{equation}\label{eq:K5}
    \begin{aligned}
       K_5^{(3)} = -\frac{4\pi^3}{3} + \pi\left[ 8L_z^2 - L_z\left(\frac{128}{z} +\frac{720}{z^2} +\frac{17984}{3 z^3} + \frac{60424}{z^4} + \frac{3400128}{5 z^5}\right) \right.\\
       \left.-\frac{32}{z} +\frac{288}{z^2} +\frac{34240}{9 z^3}+\frac{44616}{z^4} + \frac{13445568}{25 z^5} + \mathcal{O}(z^{-6})\right].
    \end{aligned}
\end{equation}
The other results are given by
\begin{equation}\label{eq:K6}
    \begin{aligned}
       K_6^{(4)} = -12\pi\zeta_3 + \pi\left[ -\frac{2L_z^3}{3} - L_z^2\left(\frac{8}{z}+\frac{36}{z^2}+\frac{512}{3 z^3}+\frac{946}{z^4}+\frac{31008}{5 z^5}\right) \right.\\
       +L_z\left(\frac{4\pi^2}{3}-\frac{24}{z}+\frac{10}{z^2}+\frac{928}{3 z^3}+\frac{16159}{6 z^4}+\frac{553976}{25 z^5}\right)\\
       +\pi^2\left(\frac{4}{3 z}+\frac{10}{z^2}+\frac{472}{9 z^3}+\frac{911}{3 z^4}+\frac{10168}{5 z^5}\right)\\
       \left.-\frac{8}{z}+\frac{57}{z^2}+\frac{5320}{27 z^3}+\frac{23233}{72 z^4}-\frac{1097524}{375 z^5} + \mathcal{O}(z^{-6})\right]. 
    \end{aligned}
\end{equation}
\begin{equation}\label{eq:K7}
    \begin{aligned}
       K_7^{(4)} = \frac{13 \pi^5}{90}+ \pi\left[\frac{L_z^4}{6} - L_z^2\left(\frac{2 \pi ^2}{3} + \frac{8}{z} + \frac{18}{z^2} + \frac{512}{9 z^3} + \frac{473}{2 z^4}+\frac{31008}{25 z^5}\right)\right.\\
        + L_z\left(12 \zeta_3 -\frac{40}{z} -\frac{13}{z^2}  +\frac{1760}{27 z^3}+\frac{13321}{24 z^4}+  \frac{98392}{25 z^5}\right) \\
        +\pi ^2 \left(\frac{4}{3 z}+\frac{5}{z^2}+\frac{472}{27 z^3}+\frac{911}{12 z^4}+\frac{10168}{25 z^5}\right)\\
        \left.-\frac{48}{z}+\frac{22}{z^2}+\frac{2360}{27 z^3}+\frac{15799}{72 z^4}+ \frac{378356}{1875 z^5} + \mathcal{O}(z^{-6})\right].
    \end{aligned}
\end{equation}

\begin{equation}\label{eq:K8}
    \begin{aligned}
       K_8^{(5)} = \frac{181 \pi ^5}{90} + \pi \left[\frac{7L_z^4}{6} + L_z^3\left(\frac{112}{3 z}-\frac{728}{3 z^2}+\frac{20608}{9 z^3}-\frac{75628}{3 z^4}+\frac{4531072}{15 z^5}\right) \right.\\
       +L_z^2\left(-\frac{13 \pi ^2}{3}+\frac{16}{z}+\frac{28}{z^2}-\frac{18848}{9 z^3} +\frac{25831}{z^4} - \frac{9139584}{25 z^5}\right)\\
       +\pi^2 L_z\left(-\frac{208}{3 z}+\frac{1352}{3 z^2}-\frac{38272}{9 z^3}+\frac{140452}{3 z^4}-\frac{8414848}{15 z^5}\right) \\
       +L_z\left(48\zeta_3-\frac{232}{z}+\frac{1309}{z^2}-\frac{318976}{27 z^3} +\frac{1046749}{8 z^4}-\frac{555427196}{375 z^5}\right)\\
       +\zeta_3\left(\frac{384}{z}-\frac{2496}{z^2}+\frac{23552}{z^3}-\frac{259296}{z^4}+\frac{15535104}{5 z^5}\right)\\
       +\pi^2\left(-\frac{40}{z}+\frac{230}{3 z^2}+\frac{1904}{9 z^3}-\frac{93895}{18 z^4}+\frac{4258208}{45 z^5}\right)\\
       \left.-\frac{400}{z}+\frac{1581}{z^2} -\frac{876064}{81 z^3}+\frac{15350453}{144 z^4}-\frac{260930249}{225 z^5}+ \mathcal{O}(z^{-6}) \right].
    \end{aligned}
\end{equation}
\begin{equation}\label{eq:K9}
    \begin{aligned}
       K_9^{(4)} = -12\pi \zeta_3 +\pi\left[ -\frac{7L_z^3}{6} - L_z^2\left(\frac{24}{z}-\frac{4}{z^2}+\frac{760}{z^3}-\frac{2914}{z^4}+\frac{283504}{5 z^5}\right)\right.\\
       +L_z\left(\frac{13 \pi ^2}{6}-\frac{66}{z}+\frac{225}{2 z^2}-\frac{148}{3 z^3}+\frac{208355}{24 z^4}-\frac{174148}{75 z^5}\right)\\
      + \pi ^2 \left(\frac{22}{3 z}-\frac{47}{3 z^2}+\frac{3004}{9 z^3}-\frac{3653}{2 z^4}+\frac{415292}{15 z^5}\right)\\
       \left.-\frac{42}{z}+\frac{509}{4 z^2}+\frac{2050}{9 z^3}+\frac{188261}{288 z^4}- \frac{3306503}{1125 z^5}+ \mathcal{O}(z^{-6}) \right].
    \end{aligned}
\end{equation}

\begin{figure}[!htp]
    \centering
    \includegraphics[width=0.5\linewidth]{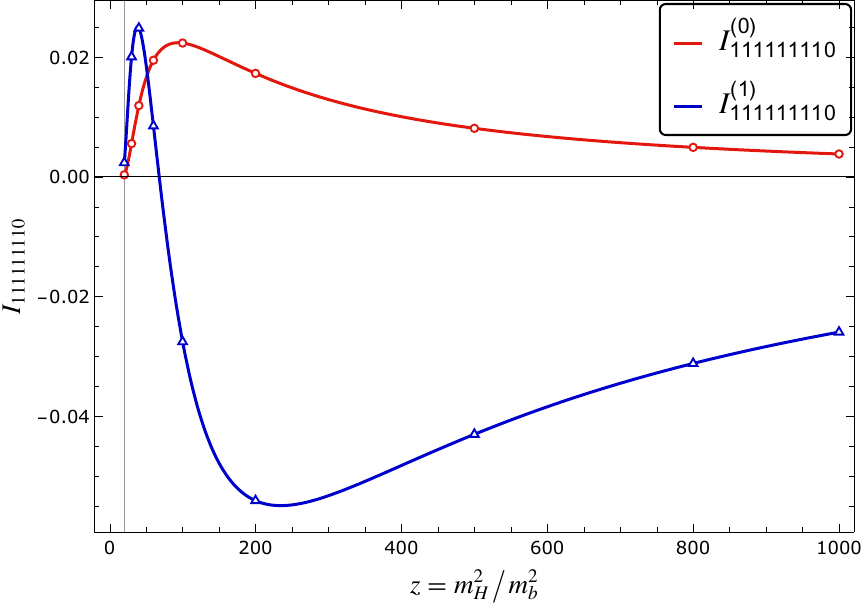}
    \caption{Comparison between analytical and numerical results of $I_{111111110}$. The solid lines are obtained from our results, while the discrete points are those from \texttt{AMFLOW}. The gray vertical line corresponds $z=20$. }
    \label{fig:I8}
\end{figure}

It is not obvious to achieve a compact form of the above results as in the banana sector. 
We leave it to future work. 
To cross-check the results of $K_8^{(5)}$ and $K_8^{(6)}$ obtained in this paper, in Fig.~\ref{fig:I8}, we show the comparison of $I_{111111110}=I_{111111110}^{(0)}+\varepsilon \cdot I^{(1)}_{111111110} + \mathcal{O}(\varepsilon^2)$ between our analytic results and the numerical evaluations obtained by \texttt{AMFLOW}. Full agreement was found. The rotation from $K_8^{(5)}$ and $K_8^{(6)}$ to $I_{111111110}^{(0)}$ and $I_{111111110}^{(1)}$ can be found in the auxiliary file. The result of $I_{111111110}^{(0)}$ also agrees with the two-dimensional numerical integration in our previous work \cite{Wang:2026zfy}.

\section{Conclusion}
\label{sec:conclusions}
In this paper, we explicitly show how to derive the $\varepsilon$-factorized differential equation of an essential three-loop integral family contributing to the decay width of $H\to b\bar{b}b\bar{b}$, and then investigate the solution space. This is a non-trivial application of the method proposed in \cite{e-collaboration:2025frv,Bree:2025tug} by taking mixing between different sectors into account. 
The results of all master integrals to any order of $\varepsilon$ can be obtained systematically from the $\varepsilon$-factorized differential equation equipped with boundary constants determined by \texttt{AMFlow}. We present results of all master integrals up to the first two orders of $\varepsilon$, despite the fact that only the leading order contributes to the decay width. 
We have described the derivation in detail, which provides a concrete example to exploit the method \cite{e-collaboration:2025frv,Bree:2025tug} in other multi-loop calculations.

\section*{Acknowledgement}
This work was supported by the National Natural Science Foundation of China under Nos. 12405117, 12321005, 12375076, 12535006. X.W. is also supported by the University Development Fund of The Chinese University of Hong Kong, Shenzhen, under the Grant No. UDF01003912.

{\footnotesize
\bibliography{refs.bib}
\bibliographystyle{h-physrev5}
}

\end{document}